\documentclass{emulateapj}
\usepackage{apjfonts}
\usepackage{CJK}

\usepackage{lscape}
\usepackage{graphicx}
\usepackage{natbib}
\usepackage{longtable}
\usepackage{rotating}

\newcommand{\etal}{et al.}
\newcommand{\hbeta}{H{$\beta$}}

\newcommand{\CIV}{C{\sevenrm IV}}

\newcommand{\MgII}{Mg{\sevenrm II}}

 \font\sevenrm=cmr7 scaled 1000

\begin{document}
\begin{CJK*}{GB}{gbsn}

\title{The Demographics of Broad-Line Quasars in the Mass-Luminosity Plane. I. Testing FWHM-Based Virial Black Hole Masses}

\shorttitle{QUASAR DEMOGRAPHICS}


\shortauthors{SHEN \& KELLY}
\author{Yue Shen and Brandon C. Kelly\altaffilmark{1}\\
Harvard-Smithsonian Center for Astrophysics, 60 Garden Street, Cambridge, MA
02138, USA.} \altaffiltext{1}{Hubble Fellow.}

\begin{abstract}

We jointly constrain the luminosity function (LF) and black hole mass
function (BHMF) of broad-line quasars with forward Bayesian modeling in the
quasar mass-luminosity plane, based on a homogeneous sample of $\sim 58,000$
SDSS DR7 quasars at $z\sim 0.3-5$. We take into account the selection effect
of the sample flux limit; more importantly, we deal with the statistical
scatter between true BH masses and FWHM-based single-epoch virial mass
estimates, as well as potential luminosity-dependent biases of these mass
estimates. The LF is tightly constrained in the regime sampled by SDSS, and
makes reasonable predictions when extrapolated to $\sim 3$ magnitudes
fainter. Downsizing is seen in the model LF. On the other hand, we find it
difficult to constrain the BHMF to within a factor of a few at $z\ga 0.7$
(with \MgII\ and \CIV-based virial BH masses). This is mainly driven by the
unknown luminosity-dependent bias of these mass estimators and its degeneracy
with other model parameters, and secondly driven by the fact that SDSS
quasars only sample the tip of the active BH population at high redshift.
Nevertheless, the most likely models favor a positive luminosity-dependent
bias for \MgII\ and possibly for \CIV, such that at fixed true BH mass,
objects with higher-than-average luminosities have over-estimated FWHM-based
virial masses. There is tentative evidence that downsizing also manifests
itself in the active BHMF, and the BH mass density in broad-line quasars
contributes an insignificant amount to the total BH mass density at all
times. Within our model uncertainties, we do not find a strong BH mass
dependence of the mean Eddington ratio; but there is evidence that the mean
Eddington ratio (at fixed BH mass) increases with redshift.

\end{abstract}

\section{Introduction}\label{sec:intro}

One major effort in modern galaxy formation studies is to understand the
cosmic evolution of supermassive black holes (SMBHs), given their ubiquitous
existence in almost every local bulge-dominant galaxy, and possible roles
during their co-evolution with the host galaxy
\citep[e.g.,][]{Magorrian_etal_1998,Gebhardt_etal_2000a,Ferrarese_Merritt_2000,Gultekin_etal_2009,
Hopkins_etal_2008a,Somerville_etal_2008}. Over the past decade, the rapidly
growing body of observational data and numerical simulations have led to a
coherent picture of the cosmic formation and evolution of SMBHs within the
hierarchical $\Lambda$CDM paradigm
\citep[e.g.,][]{Haiman_Loeb_1998,Kauffmann_Haehnelt_2000,Wyithe_Loeb_2003,Volonteri_etal_2003,Hopkins_etal_2006,
Hopkins_etal_2008a,Shankar_etal_2009a,Shankar_etal_2010,Shen_2009}. Although
many fundamental issues regarding SMBH growth still remain unclear (such as BH seeds,
fueling and feedback mechanisms), these cosmological SMBH models are starting
to reproduce a variety of observed SMBH statistics in an unprecedented
manner.

It is now widely appreciated that SMBHs grow by gas accretion in the past,
during which they are witnessed as quasars and active galactic nuclei (AGNs)
\citep[e.g.,][]{Salpeter_1964,Zeldovich_Novikov_1964,Lynden-Bell_1969}. In
the local Universe, the mass function of dormant SMBHs is estimated by
convolving the galaxy distribution functions with various scaling relations
between galaxy properties and BH mass. This relic SMBH population has been
used to constrain the accretion history of their active counterparts, using
the So{\l}tan argument and its extensions
\citep[e.g.,][]{Soltan_1982,Small_Blandford_1992,Salucci_etal_1999,Yu_Tremaine_2002,Yu_Lu_2004,Yu_Lu_2008,Shankar_etal_2004,Shankar_etal_2009a,
Marconi_etal_2004,Merloni_2004,Hopkins_etal_2007a,Merloni_Heinz_2008}. The
agreement between the relic BH mass density and the accreted mass density
provides compelling evidence that these two populations are ultimately
connected. Therefore it is of imminent importance to quantify the abundance
of active SMBHs as a function of redshift.

The demographics of the active SMBH population has been the central topic for
quasar studies since the first discovery of quasars
\citep[][]{Schmidt_1963,Schmidt_1968}. Traditionally this is done in terms of
the luminosity function (LF), i.e., the abundance of objects at different
luminosities. Measuring the LF and its evolution has been the most important
goal for modern quasar surveys
\citep[e.g.,][]{Schmidt_Green_1983,Green_etal_1986}. In the last decade the
LF has been measured for different populations of active SMBHs and in
different bands \citep[e.g.,][]{Fan_etal_2001,Fan_etal_2004,Boyle_etal_2000,
Wolf_etal_2003,Croom_etal_2004,Croom_etal_2009,Hao_etal_2005b,
Richards_etal_2005,Richards_etal_2006a,Jiang_etal_2006,Jiang_etal_2008,Jiang_etal_2009,
Fontanot_etal_2007,Bongiorno_etal_2007, Willott_etal_2010b,
Ueda_etal_2003,Hasinger_etal_2005,Silverman_etal_2005,Silverman_etal_2008,Barger_etal_2005},
and it constitutes a crucial observational component in all cosmological SMBH
models.

A more important physical quantity of SMBHs is BH mass. BH mass is directly
related to growth, and when the BH is active, it determines the accretion
efficiency ($\dot{M}/M_{\rm BH}$) via the Eddington ratio and an assumed
radiative efficiency (e.g., such as the average value constrained by the
Soltan argument). Thus knowing the mass function of SMBHs as a function of
redshift adds significantly to our understandings of their cosmic evolution.

It remains challenging to directly measure the dormant BHMF at high redshift.
This is not only because the galaxy distribution functions are less
well-constrained at high redshift, but also because the evolution of the
scaling relations (both the mean relation and the scatter) between galaxy
properties and BH mass is poorly understood. On the other hand, it has become
possible to measure the active BHMF of broad-line quasars\footnote{From now
on, unless otherwise specified, we use the term ``quasar'' to refer to
broad-line (type 1) quasars for simplicity.}, using the so-called virial BH
mass estimators based on their broad emission line and continuum properties
measured from single-epoch spectra
\citep[e.g.,][]{Wandel_etal_1999,Mclure_Dunlop_2004,Vestergaard_Peterson_2006},
a technique rooted on reverberation mapping (RM) studies of local broad-line
AGNs
\citep[e.g.,][]{Blandford_McKee_1982,Peterson_1993,Kaspi_etal_2000,Peterson_etal_2004,
Bentz_etal_2006,Bentz_etal_2009}. These single-epoch virial BH mass
estimators are calibrated empirically using the RM AGN sample to yield on
average consistent BH mass estimates compared with RM masses, which are
further tied to the BH masses predicted using the $M_{\rm BH}-\sigma$
relation \citep[e.g.,][]{Tremaine_etal_2002,Onken_etal_2004}. The nominal
scatter between these single-epoch virial estimates and the RM masses is on
the order of $\sim 0.4$ dex
\citep[e.g.,][]{Mclure_Jarvis_2002,Mclure_Dunlop_2004,Vestergaard_Peterson_2006}.

A couple of recent studies have applied this technique to measure the active
BHMF with statistical quasar and AGN samples
\citep[e.g.][]{Greene_Ho_2007a,Vestergaard_etal_2008,Vestergaard_Osmer_2009,
Schulze_Wisotzki_2010}. A robust determination of the active BHMF constitutes
an important building block of cosmological SMBH models, in addition to the
luminosity function. These virial mass estimators also enable statistical
studies on the Eddington ratios of broad-line quasars and AGNs
\citep[e.g.,][]{Vestergaard_2004,Mclure_Dunlop_2004,Kollmeier_etal_2006,Sulentic_etal_2006,Babic_etal_2007,
Jiang_etal_2007,Kurk_etal_2007,Netzer_etal_2007,Shen_etal_2008b,Gavignaud_etal_2008,
Labita_etal_2009a,Trump_etal_2009,Trump_etal_2011,Willott_etal_2010,Trakhtenbrot_etal_2011},
over a wide range of luminosities and redshifts, and therefore provide
constraints on the accretion efficiency of these active SMBHs.

With the development of these virial mass estimators, we now have both BH
mass estimates and luminosities for broad-line quasar samples. Given the
intimate relation between BH mass and luminosity, it is important and
necessary to study their joint distribution and evolution in the
mass-luminosity plane
\citep[e.g.,][]{Steinhardt_Elvis_2010a,Steinhardt_etal_2011a}. This
represents a significant step forward to study the demography of quasars than
using LF alone, and offers new insights on the properties and evolution of
the active SMBH population.

However, the importance of distinguishing between virial mass estimates and
true BH masses can hardly be over-stressed. While these virial estimators
currently are the only practical way to estimate BH masses for large samples
of broad-line quasars and AGNs, the nontrivial uncertainty of these imperfect
estimators has severe impact on the mass distribution under study. The
difference between virial masses and true masses not only modifies the
underlying true distribution, but also introduces Malmquist-type biases
\citep[e.g.,][]{Shen_etal_2008b,Kelly_etal_2009a,Shen_Kelly_2010}. These
effects tend to dilute any potential mass-dependent trends or correlations
\citep[e.g.,][]{Kelly_Bechtold_2007,Shen_etal_2009}, and may lead to
unreliable conclusions. Thus it is important to consider these effects when
the statistics is becoming good enough.

\citet{Kelly_etal_2009a} developed a Bayesian framework to estimate the
BHMF/LF for broad-line quasars, which accounts for the uncertainty in virial
BH mass estimates, as well as the selection incompleteness in BH mass (since
the sample is selected in luminosity). This method was subsequently applied
to the SDSS DR3 quasar sample \citep{Kelly_etal_2010}, based on virial mass
estimates from \citet{Vestergaard_etal_2008}. This Bayesian framework is a
more rigorous and quantitative treatment than the simple forward modeling
performed in \citet{Shen_etal_2008b}, and allows a more reliable measurement
of the true active BHMF and its uncertainty for quasars.

Equipped with an improved version of this Bayesian framework, in this paper
we measure the active BHMF and LF based on a homogeneous sample of $\sim
58,000$ quasars from SDSS DR7 with FWHM-based virial mass estimates from
\citet{Shen_etal_2011a}. The much improved statistics now allows a detailed
examination of the joint distribution in the mass-luminosity plane, and
provides better constraints on BH accretion properties.

A key difference in our approach compared with most earlier work is the
attempt to account for the uncertainty (error) in these virial mass
estimates. We distinguish three types of errors in single-epoch virial BH
mass estimates:
\begin{itemize}

\item  measurement error, which is propagated from the uncertainties of
    FWHM and continuum luminosity measurements from the spectra; the
    measurement errors are typically $\ll 0.3$ dex for our sample (see
    Fig.\ \ref{fig:bh_dist}) and hence are negligible; however,
    measurement error may become important for other samples with low
    spectral quality.

\item statistical error, which is the scatter of virial BH masses around
    RM masses when these virial estimators were calibrated against local
    RM AGN sample; the statistical error is $\ga 0.3$ dex
    \citep[e.g.,][]{Mclure_Jarvis_2002,
    Mclure_Dunlop_2004,Vestergaard_Peterson_2006}, which will be taken
    into account in our Bayesian approach.

\item  systematic biases, which may result from the virial assumption,
    the usage of RM masses as true masses during calibration, the usage
    of a particular definition of line width as the surrogate for the
    virial velocity, the extrapolation of the virial calibrations to high
    luminosity/redhift, as well as other possible systematics
    \citep[e.g.,][]{Krolik_2001,Collin_etal_2006,Shen_etal_2008b,Marconi_etal_2008,
    Fine_etal_2008, Fine_etal_2010, Netzer_2009,
    Denney_etal_2009,Wang_etal_2009b,Graham_etal_2011,
    Rafiee_Hall_2011b,Steinhardt_2011}.

    We generally neglect systematic biases in the current study, as they
    are poorly understood at present. That means we assume {\em on
    average} these virial mass estimators give unbiased mass estimates
    (see \S\ref{subsec:prelim} for the meaning of ``unbiased''). However,
    we do consider a possible luminosity-dependent bias
    \citep[e.g.,][]{Shen_etal_2008b,Shen_Kelly_2010}, which we describe
    in detail in \S\ref{subsec:prelim}. This is not only because
    luminosity is an explicit term in all virial estimators, but also
    because that many studies with virial BH masses are restricted to
    finite luminosity bins or flux-limited samples. Moreover,
    understanding any potential luminosity-dependent bias is crucial to
    probe the true distribution in the mass-luminosity plane.

\end{itemize}

This paper is organized as follows. In \S\ref{sec:data} we describe the data;
we present the traditional binned LF/BHMF in \S\ref{subsec:bin_BHMF} and
describe the Bayesian approach in \S\ref{subsec:MCMC_BHMF}. We present our
model results in \S\ref{sec:results}, discuss the results in \S\ref{sec:disc}
and conclude in \S\ref{sec:con}. Throughout the paper we adopt a flat
$\Lambda$CDM cosmology with cosmological parameters $\Omega_{\Lambda}=0.7$,
$\Omega_0=0.3$, $h=0.7$, to match most of the recent quasar demographics
work. Volume is in comoving units unless otherwise stated. We distinguish
virial masses from true masses with a subscript $_{\rm vir}$ or $_e$. Quasar
luminosity is expressed in terms of the rest-frame 2500\,\AA\ continuum
luminosity ($L\equiv \lambda L_\lambda$ or $l\equiv \log L$ for short), and
we adopt a constant bolometric correction $C_{\rm bol}=L_{\rm bol}/L=5$.

\section{The Data}\label{sec:data}

\begin{figure*}
  \centering
    \includegraphics[width=0.48\textwidth]{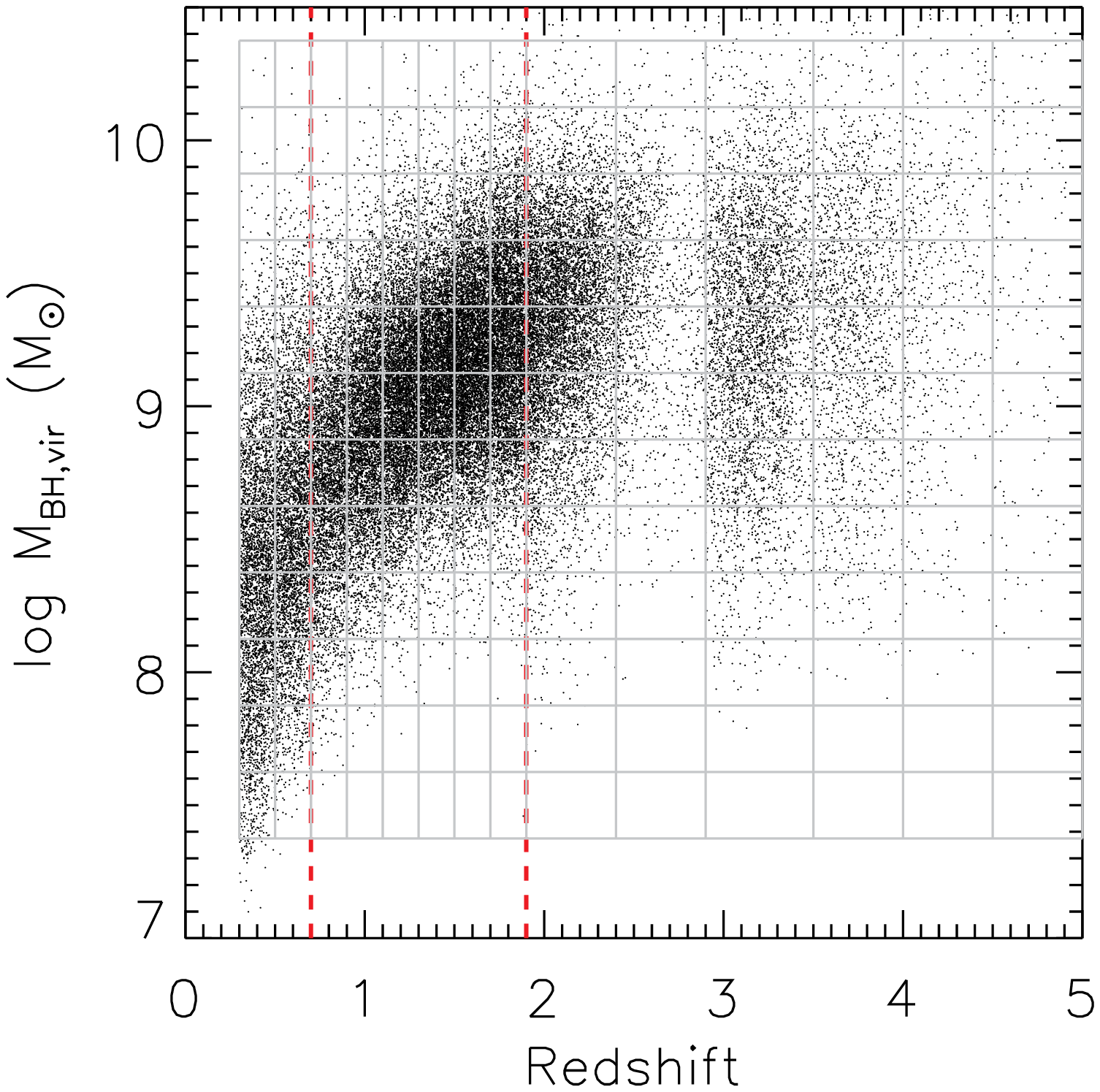}
    \includegraphics[width=0.48\textwidth]{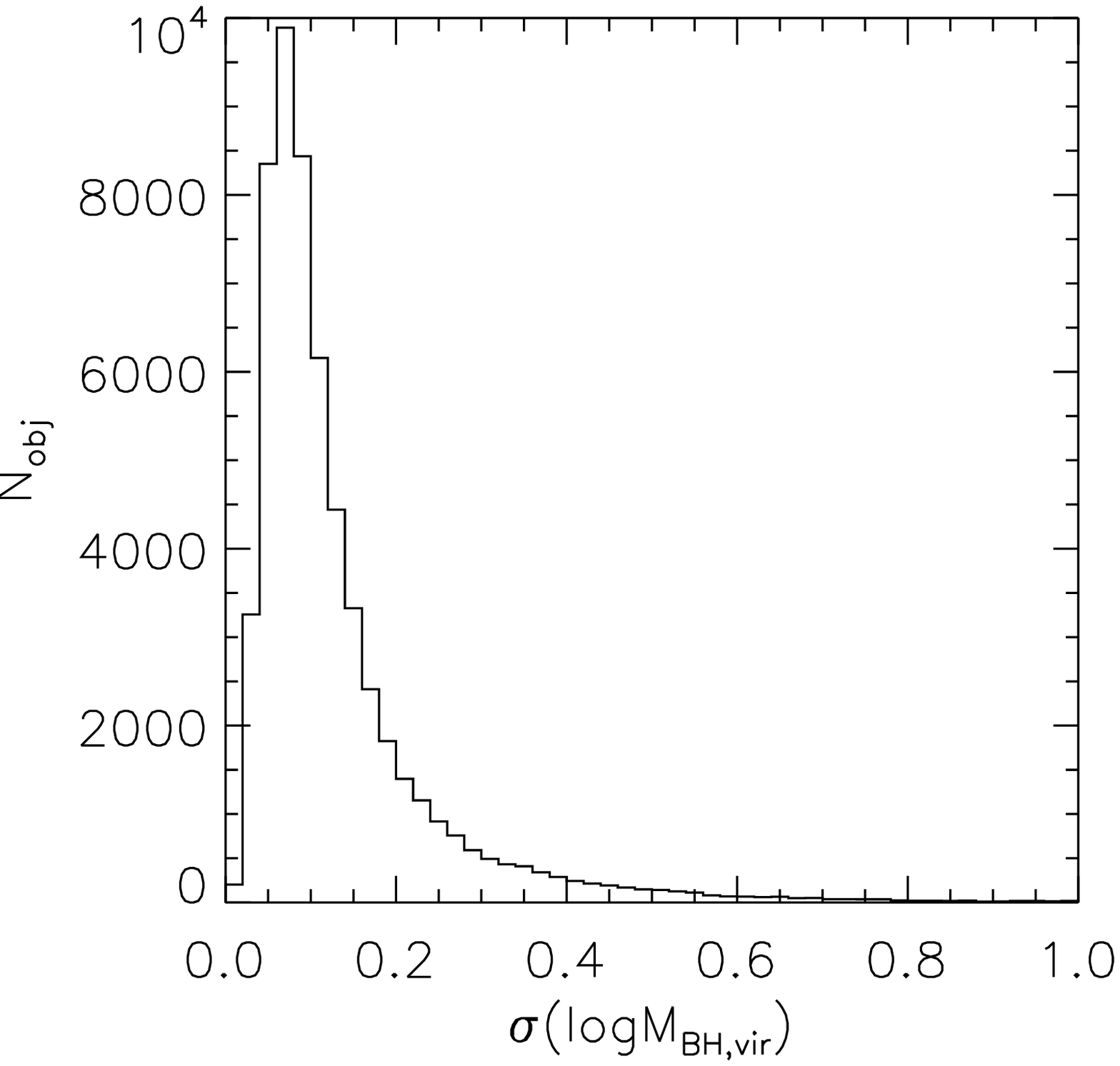}
    \caption{{\em Left}: Redshift distribution of virial BH masses in our sample. {\em Right}: Distribution of measurement errors of the virial BH
    mass estimates. The vast majority of virial mass estimates have negligible measurement errors compared with the nominal statistical uncertainty
    of
    virial BH mass estimators $\sigma_{\rm vir}\sim 0.3-0.4$ dex.}
    \label{fig:bh_dist}
\end{figure*}

Our parent sample is the SDSS DR7 quasar catalog \citep{Schneider_etal_2010},
which contains 105,783 bona fide quasars with $i$-band absolute magnitude
$M_i<-22$ and have at least one broad emission line (${\rm FWHM}>1000 {\rm\
km\ s^{-1}}$) or have interesting/complex absorption features. Among these
quasars, about half were targeted using the final quasar target algorithm
described in \citet{Richards_etal_2002a}, and form a homogeneous, statistical
quasar sample \citep[e.g.,][]{Richards_etal_2006a,Shen_etal_2007b}, which we
adopt in the current study. Quasars in this homogeneous sample are
flux-limited to $i=19.1$ below $z=2.9$ and to $i=20.2$ beyond\footnote{There
are a tiny fraction of uniformly-selected quasars targeted by the {\tt HiZ}
branch of the target selection algorithm \citep{Richards_etal_2002a} at
$z<2.9$ down to $i=20.2$. We have rejected these quasars in our flux-limited
sample \citep[see ][for more details]{Shen_etal_2011a}.}. There is also a
bright limit of $i=15$ for SDSS quasar targets, which only becomes important
for the most luminous quasars at the lowest redshift \citep[see Fig.\ 1
in][]{Shen_etal_2011a}. We have used the continuum and emission line
$K$-corrections in \citet{Richards_etal_2006a} to compute the absolute
$i$-band magnitude normalized at $z=2$, $M_i[z=2]$. At $z<0.5$, host
contamination becomes more and more prominent towards lower redshift
\citep[e.g.,][]{Shen_etal_2011a}, so we restrict our sample to $z\ge 0.3$.
Our final sample includes 57,959 quasars at $0.3\le z\le 5$. The sky coverage
of this uniform quasar sample is carefully determined, using the approach
detailed in the appendix in \citet{Shen_etal_2007b}, to be 6248 ${\rm
deg}^2$.

The virial mass estimates and measurement errors for these quasars were taken
from \citet{Shen_etal_2011a}. We refer the reader to \citet{Shen_etal_2011a}
for details regarding the spectral measurements and virial mass estimates. In
short, the spectral region around each of the three lines (\hbeta, \MgII, and
\CIV) is fit by a power-law continuum plus iron template\footnote{Except for
\CIV, where we only fit a power-law continuum with no iron template
applied.}, and a set of Gaussians for the line emission. Narrow line emission
is modeled for \hbeta\ and \MgII\ but not for \CIV. We use the continuum
luminosity and line FWHM from the spectral fits to compute a virial mass
using one of the fiducial virial calibrations adopted in \citet[][eqns.
5,6,8]{Shen_etal_2011a}. $>95\%$ of the 57,959 quasars have measurable virial
BH masses. Fig.\ \ref{fig:bh_dist} (right) shows the distribution of
measurement errors (propagated from the FWHM and continuum luminosity errors)
of these virial mass estimates. The vast majority of virial estimates have a
measurement error far below $0.3-0.4$ dex, the nominal statistical
uncertainty of virial estimators. Three line estimators were used: \hbeta\
\citep[][$z<0.7$]{Vestergaard_Peterson_2006}; \MgII\ \citep[][$0.7\le
z<1.9$]{Shen_etal_2011a}; \CIV\ \citep[][$z>1.9$]{Vestergaard_Peterson_2006}.
Virial BH masses based on two estimators are smoothly bridged across the
dividing redshift, i.e., there is no systematic offset between two different
estimators. Fig.\ \ref{fig:bh_dist} (left) shows the redshift distribution of
virial mass estimates in our sample, where the vertical dashed lines mark the
divisions between two estimators and the grid we use to compute the BHMF (see
below) is shown in gray. We reject objects with a measurement error $>0.5$
dex in virial mass estimates in computing the BHMF, and we will correct for
this incompleteness in mass estimates in Sec \ref{sec:BHMF}.

\section{The Quasar LF and BHMF}\label{sec:BHMF}

\subsection{The Traditional Approach}\label{subsec:bin_BHMF}

\begin{figure}
  \centering
    \includegraphics[width=0.45\textwidth]{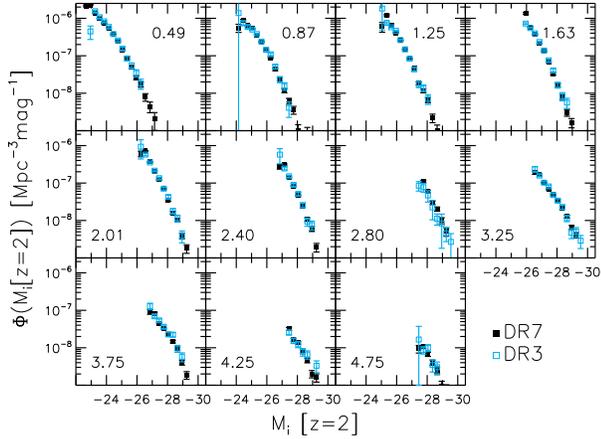}
    \caption{Comparison between the DR3 \citep{Richards_etal_2006a} and DR7 binned LF (this work)
    for the same luminosity-redshift grid. The DR7 results are in good agreement with earlier
    DR3 results. }
    \label{fig:lf_dr7_dr3}
\end{figure}

\begin{deluxetable}{lccc}
\tablecaption{Summary of \texttt{zbins} \label{table:zbin}} \tablehead{ zbin
& $z$ range & $N_{Q}/N_{\rm vir}$ & $\bar{M}_{i,{\rm lim}}[z=2]$} \startdata
\hbeta\ & & & \\
1\dotfill & $[0.3,0.5]$ & 4298/4149 &  $-22.94$ \\
2\dotfill & $[0.5,0.7]$ & 4206/4027 &  $-23.84$ \\
\MgII\ & & & \\
3\dotfill & $[0.7,0.9]$ & 3955/3873 & $-24.61$ \\
4\dotfill & $[0.9,1.1]$ & 4871/4772 & $-25.06$ \\
5\dotfill & $[1.1,1.3]$ & 5872/5789 & $-25.39$ \\
6\dotfill & $[1.3,1.5]$ & 5925/5855 & $-25.73$ \\
7\dotfill & $[1.5,1.7]$ & 6459/6340 & $-25.99$ \\
8\dotfill & $[1.7,1.9]$ & 5839/5566 & $-26.29$ \\
\CIV\ & & & \\
9\dotfill  & $[1.9,2.4]$ & 7761/7545 & $-26.83$ \\
10\dotfill & $[2.4,2.9]$ & 1695/1641 & $-27.34$ \\
11\dotfill & $[2.9,3.5]$ & 4317/4003 & $-26.66$ \\
12\dotfill & $[3.5,4.0]$ & 1830/1666 & $-27.00$ \\
13\dotfill & $[4.0,4.5]$ & 661/518 & $-27.36$ \\
14\dotfill & $[4.5,5.0]$ & 270/152 & $-27.45$ \\
\enddata
\tablecomments{The second column lists the boundaries of each \texttt{zbin}.
The third column lists the total number of quasars and those with measurable
virial masses (measurement error $<0.5$ dex) in each \texttt{zbin}. The
fourth column lists the limiting luminosity in terms of the absolute $i$-band
magnitude normalized at $z=2$ \citep[][]{Richards_etal_2006a}, which
corresponds to the flux limit ($i=19.1$ and 20.2 for $z<2.9$ and $z>2.9$) and
is estimated at the median redshift for each \texttt{zbin}.}
\end{deluxetable}

\begin{deluxetable}{lccc}
\tablecaption{Binned DR7 virial BHMF \label{table:phi_mvir}} \tablehead{
$\bar{z}$ & $\log M_{\rm BH,vir}$ & $\log\Phi(M_{\rm BH,vir})$ & $\log\sigma(M_{\rm BH,vir})$\\
& $(M_\odot)$ & (Mpc$^{-3}\log M_{\rm BH,vir}^{-1}$) & (Mpc$^{-3}\log M_{\rm
BH,vir}^{-1}$) } \startdata
0.4 & 7.50 & $-6.378$  & $-7.370$ \\
0.4 & 7.75 & $-5.957$  & $-7.156$ \\
0.4 & 8.00 & $-5.813$  & $-7.110$ \\
\enddata
\tablecomments{The full table is available in the electronic version of the
paper. }
\end{deluxetable}

\begin{deluxetable}{lccc}
\tablecaption{Binned DR7 LF \label{table:phi_l}} \tablehead{
$\bar{z}$ & $M_{i}[z=2]$ & $\log\Phi(M_i[z=2])$ & $\log\sigma(M_i[z=2])$\\
& & (Mpc$^{-3}{\rm mag}^{-1}$) & (Mpc$^{-3}{\rm mag}^{-1}$) } \startdata
0.4 & $-22.65$  & $-5.669$ & $-6.920$ \\
0.4 & $-22.95$  & $-5.643$ & $-7.078$ \\
0.4 & $-23.25$  & $-5.858$ & $-7.350$ \\
\enddata
\tablecomments{The full table is available in the electronic version of the
paper. }
\end{deluxetable}

Following the common practice in the literature
\citep[e.g.,][]{Fan_etal_2001,Richards_etal_2006a,Greene_Ho_2007a,Vestergaard_etal_2008,
Vestergaard_Osmer_2009,Schulze_Wisotzki_2010}, we use the $1/V_{\rm max}$
method \citep[e.g.,][]{Schmidt_1968} to estimate the LF and active BHMF:
\begin{equation}\label{eqn:vmax}
V_{\rm max}=\frac{\Omega}{4\pi}\int_{\rm z_{min}}^{\rm z_{max}}\Theta(L,z)\frac{dV_c}{dz}dz\ ,
\end{equation}
where $\Omega$ is the sky coverage of our sample, $dV_c/dz$ is the
differential comoving volume, ${\rm z_{min}}$ and ${\rm z_{max}}$ are the
minimum and maximum redshift within a redshift-luminosity (virial mass) bin
that is accessible for a quasar with luminosity $L$, and $\Theta(L,z)$ is the
{\em luminosity} selection function mapped on a two-dimensional grid of
luminosity and redshift. We use the tabulated selection
function\footnote{There is no difference in the target selection completeness
between the uniform DR3 quasars used in \citet{Richards_etal_2006a} and the
uniform DR7 quasars used here, since the final quasar target algorithm was
implemented after DR1.} in \citet{Richards_etal_2006a} with interpolation to
estimate $\Theta(L,z)$, and calculate $V_{\rm max}$ for each quasar in a
redshift-luminosity (virial mass) bin. The binned LF, $\Phi(M_i[z=2])\equiv
dn/dM_i[z=2]$ is then:
\begin{equation}
\Phi(M_i[z=2])=\frac{1}{\Delta M_i[z=2]}\sum_{j=1}^N\left(\frac{1}{V_{\rm max,j}}\right)\ ,
\end{equation}
with a Poisson statistical uncertainty
\begin{equation}
\sigma(\Phi)=\frac{1}{\Delta M_i[z=2]}\bigg[\sum_{j=1}^{N} \bigg(\frac{1}{V_{{\rm max},j}}\bigg)^2\bigg]^{1/2}\ ,
\end{equation}
where the summation is over all quasars within a redshift-magnitude bin.

The $1/V_{\rm max}$ binned BHMF, $\Phi(M_{\rm BH,vir})\equiv dn/d\log M_{\rm BH,vir}$, is then:
\begin{equation}
\Phi(M_{\rm BH,vir})=\frac{1}{\Delta\log M_{\rm BH,vir}}\sum_{j=1}^{N} \bigg(\frac{1}{V_{{\rm max},j}}\bigg)\ ,
\end{equation}
with a Poisson statistical uncertainty
\begin{equation}
\sigma(\Phi)=\frac{1}{\Delta\log M_{\rm BH,vir}}\bigg[\sum_{j=1}^{N} \bigg(\frac{1}{V_{{\rm max},j}}\bigg)^2\bigg]^{1/2}\ ,
\end{equation}
where the summation is over all quasars within a redshift-mass bin.

As a sanity check, we computed the DR7 quasar luminosity function (LF) in the
same $L-z$ grid as in \citet{Richards_etal_2006a}, and found it in excellent
agreement with the DR3 results with smaller statistical error bars (Fig.\
\ref{fig:lf_dr7_dr3}).

\begin{figure*}
  \centering
    \includegraphics[width=0.9\textwidth]{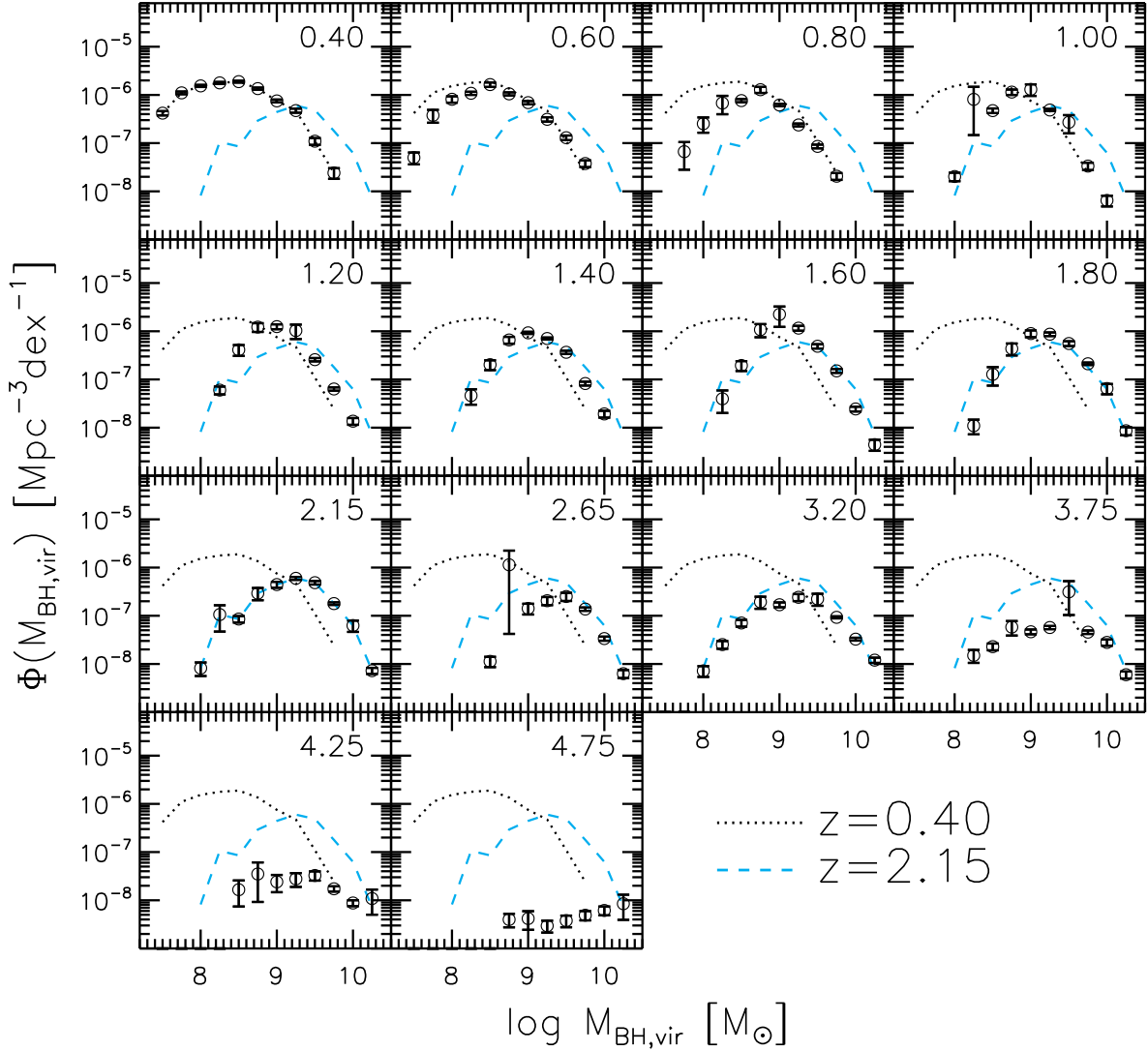}
    \caption{Binned virial BHMF using the $1/V_{\rm max}$ technique. In each panel the
    points with error bars are the results for each \texttt{zbin}, and the dotted
    and dashed lines are reference results in \texttt{zbin1} and \texttt{zbin9}.
    The mean redshift in each \texttt{zbin} is marked on the upper-right corner of
    each panel. }
    \label{fig:vmax_bhmf_dlogm}
\end{figure*}

To compute the binned LF and BHMF we choose a redshift grid (\texttt{zbins})
that avoids straddling two mass estimators, with boundaries of 0.3, 0.5, 0.7,
0.9, 1.1, 1.3, 1.5, 1.7, 1.9, 2.4, 2.9, 3.5, 4.0, 4.5, 5.0. Within each of
the 14 \texttt{zbins} we use a mass grid with a bin size of $\Delta\log
M_{\rm BH,vir}=0.25$ starting from $\log M_{\rm BH,vir}=7.375$, and a
luminosity grid with a bin size of $\Delta M_i[z=2]=0.3$ starting from
$M_i[z=2]=-22.5$. Table \ref{table:zbin} summarizes information for each
\texttt{zbin}. Fig.\ \ref{fig:vmax_bhmf_dlogm} shows the binned virial BHMF
using the $1/V_{\rm max}$ technique. Our binned virial BHMF results are
similar to the binned virial BHMF estimated in \citet{Vestergaard_etal_2008}
based on DR3 quasars, with much better statistics due to increased sample
size.

Two important facts limit the application of the binned virial BHMF. First,
it is inappropriate to use the selection function upon luminosity selection
for the BHMF, i.e., BHs with instantaneous luminosity fainter than the flux
limit of the survey will be missed regardless of their masses. As a result,
the binned BHMF suffers from incompleteness, especially at the low-mass end,
and the turn-over of the BHMF at low masses seen in Fig.\
\ref{fig:vmax_bhmf_dlogm} is not real. Second, virial BH masses are {\em not}
true masses. Substantial scatter between virial mass estimates and the true
masses changes the underlying BH mass distribution, and may lead to
significant Malmquist-type biases
\citep[e.g.,][]{Shen_etal_2008b,Kelly_etal_2009a,Kelly_etal_2010,Shen_Kelly_2010}.
The latter effect is particularly important at the high-mass end (where the
contamination from intrinsically lighter BHs can dominate over the indigenous
population) and at high redshift \citep[where the virial BH mass estimator switches
to the more problematic \CIV\ line, e.g.,][]{Shen_etal_2008b}. The Bayesian framework developed in
\citet{Kelly_etal_2009a} and described in \S\ref{subsec:MCMC_BHMF} remedies
these issues, and provides more reliable estimates for the intrinsic BHMF.

Bearing in mind the limitations of the binned BHMF, Fig.\
\ref{fig:vmax_bhmf_dlogm} shows a coherent evolution for the most massive
($M_{\rm BH,vir}\ga 3\times 10^9\ M_\odot$) BHs: their abundance rises from
high redshift and reaches maximum around $z\sim 2$, then decreases towards
lower redshift. This trend is likely a manifestation of the rise and fall of
bright quasars seen in the LF, and we will test this trend with the Bayesian
approach described in Sec \ref{subsec:MCMC_BHMF}.

For future comparison purposes only, we tabulated the binned virial BHMF in
Table \ref{table:phi_mvir}; but we remind the reader that it should be
interpreted with caution. We also tabulated the binned LF in Table
\ref{table:phi_l}.

\tabletypesize{\tiny}
\begin{deluxetable*}{lccccccccccccccccccc}
\tablecaption{Model LF, BHMF and Eddington ratio
function \label{table:phi_m_model}} \tablehead{ & & &
\multicolumn{3}{c}{$\log\Phi(L)$} & & \multicolumn{3}{c}{$\log\Phi(M_{\rm
BH})$} & \multicolumn{3}{c}{$\log\Phi(M_{\rm BH,det})$} & &
\multicolumn{3}{c}{$\log\Phi(\lambda)$} &
\multicolumn{3}{c}{$\log\Phi(\lambda_{\rm det})$} \\
\hline\\ $\bar{z}$ & $M_i[z=2]$ & $\log L$ & $\Phi_0$ & $\Phi_+$ & $\Phi_{-}$
&  $\log M_{\rm BH}$ & $\Phi_0$ & $\Phi_{+}$ & $\Phi_{-}$
& $\Phi_0$  & $\Phi_{+}$ & $\Phi_{-}$& $\log\lambda$ & $\Phi_0$ & $\Phi_{+}$& $\Phi_{-}$ & $\Phi_0$ & $\Phi_+$ &  $\Phi_-$ \\
& & $({\rm erg\,s^{-1}})$ & \multicolumn{3}{c}{(Mpc$^{-3}{\rm dex}^{-1}$)} &
$(M_\odot)$ & \multicolumn{3}{c}{(Mpc$^{-3}{\rm dex}^{-1}$)} &
\multicolumn{3}{c}{(Mpc$^{-3}{\rm dex}^{-1}$)} & &
\multicolumn{3}{c}{(Mpc$^{-3}{\rm dex}^{-1}$)} &
\multicolumn{3}{c}{(Mpc$^{-3}{\rm dex}^{-1}$)} } \startdata
0.4 & $-16.904$ & $42.00$ & $-6.299$ & $-6.002$ & $-6.639$ & $6.000$ & $-10.684$ & $-8.179$ & $-12.383$ & $-19.804$ & $-17.507$ & $-21.693$ &
$-4.000$
& $-9.482$ & $-9.108$ & $-10.225$ & $-17.622$ & $-16.349$ & $-19.266$ \\
0.4 & $-16.979$ & $42.03$ & $-6.240$ & $-5.950$ & $-6.572$ & $6.025$ & $-10.550$ & $-8.114$ & $-12.201$ & $-19.521$ & $-17.287$ & $-21.363$ &
$-3.975$
& $-9.368$ & $-9.002$ & $-10.097$ & $-17.397$ & $-16.152$ & $-18.999$ \\
0.4 & $-17.054$ & $42.06$ & $-6.182$ & $-5.900$ & $-6.506$ & $6.050$ & $-10.415$ & $-8.049$ & $-12.023$ & $-19.240$ & $-17.068$ & $-21.038$ &
$-3.950$
& $-9.256$ & $-8.896$ & $-9.970$  & $-17.175$ & $-15.955$ & $-18.736$ \\
\enddata
\tablecomments{The full table is available in the electronic version of the
paper. }
\end{deluxetable*}


\subsection{The Bayesian Approach}\label{subsec:MCMC_BHMF}

As discussed earlier, the causal connection between the LF and BHMF naturally
requires a determination of the joint distribution in the mass-luminosity
plane. In doing so, one needs to account for selection effects of the flux
limit of the sample, and to distinguish between virial masses and true
masses. The best approach is a forward modeling, in which we specify an
underlying distribution of true masses and luminosities and map to the
observed mass-luminosity plane by imposing the flux limit and relations
between virial masses and true masses, and compare with the observed
distribution
\citep[e.g.,][]{Shen_etal_2008b,Kelly_etal_2009a,Kelly_etal_2010}. This is a
complicated and model-dependent problem. Below we first demonstrate our best
understandings of the relationship between virial masses and true masses,
then we describe our model parameterizations and the implementation of the
Bayesian framework. We defer the caveats in our model to
\S\ref{sec:disc_caveats}.

\subsubsection{Preliminaries}\label{subsec:prelim}

Here we describe our modeling of the statistical errors of virial mass
estimates, under the premise that these FWHM-based virial mass estimators on
average give the correct mean (e.g., see Eqn.\ 8 below). For clarity, we use
$p(x|y)$ to denote the conditional probability distribution of quantity $x$
at fixed $y$, and $x|y$ to denote a random value of $x$ at fixed $y$ drawn
from $p(x|y)$.

For the local RM AGN sample (which has a dispersion of $\sim 1$ dex in
luminosity), single-epoch virial BH mass estimates were calibrated against RM
masses (assumed to be true masses) to have the right mean, and a scatter
(uncertainty) of $\sim 0.4$ dex around RM masses
\citep[e.g.,][]{Mclure_Jarvis_2002,Mclure_Dunlop_2004,Vestergaard_Peterson_2006}.
To account for the effects of the uncertainty in virial BH mass estimates, we
first must understand the origin of this uncertainty. It is natural to
ascribe this uncertainty to two facts
\citep[e.g.,][]{Shen_etal_2008b,Shen_Kelly_2010}: a) luminosity is an
imperfect tracer of the BLR size; b) line width is an imperfect tracer of the
virial velocity. Taken together, some portion of the variations in luminosity
and line width are independent of each other, causing the virial mass
estimates to scatter around the true value; the remaining portion of the
variations in luminosity and line width cancel with each other, and do not
contribute to the scatter in the virial mass estimates.

To better understand this, consider the following example. Take a population
of $N$ BHs with the same true mass $m\equiv \log M_{\rm BH}$, and assuming:
a) The FWHM and luminosity follow lognormal distributions at this fixed true
BH mass; b) a {\em mean} luminosity-radius ($R-L$) relation $R\propto
L^{0.5}$, and a linear {\em mean} relation between FWHM and the virial
velocity $v$; and c) the virial masses are unbiased on average. For this
population of BHs, the luminosity $l\equiv \log L$ of individual object is
given by:
\begin{equation}\label{eqn:dist_L}
l|m= \langle l\rangle_m + G_1(0|\sigma_{l}^\prime) + G_0(0|\sigma_{\rm corr})\ ,
\end{equation}
where $l|m$ is the individual object luminosity at this fixed $m$,
$G_i(\mu|\sigma)$ is a Gaussian random deviate with mean $\mu$ and dispersion
$\sigma$, and $\langle l\rangle_m$ is the expectation value of luminosity at
this true BH mass. Similarly we can generate individual line width $w\equiv
\log {\rm FWHM}$ as:
\begin{equation}\label{eqn:dist_FWHM}
w|m= \langle w\rangle_m + G_2(0|\sigma_{w}) - 0.25G_0(0|\sigma_{\rm corr})\ ,
\end{equation}
where $\langle w\rangle_m$ is the expectation value of line width at this
true BH mass. The individual virial mass estimate $m_e\equiv\log M_{\rm
BH,vir}$ at this fixed $M_{\rm BH}$ is then
\begin{equation}\label{eqn:p_me_m}
m_e|m= m + 0.5G_1(0|\sigma_{l}^\prime)+ 2G_2(0|\sigma_{w})\ ,
\end{equation}
which implies that the virial BH mass estimates follow a lognormal
distribution around the correct mean (i.e., $m$), but have a lognormal
scatter (virial uncertainty) $\sigma_{\rm vir}=\sqrt{(0.5\sigma_l^\prime)^2 +
(2\sigma_{w})^2}$ around the mean
\citep[e.g.,][]{Shen_etal_2008b,Shen_Kelly_2010}. The $G_0$ terms of
variation in luminosity and FWHM exactly cancel with each other and do not
contribute to the virial uncertainty, and were referred to as the
``correlated variations'' in FWHM and luminosity in the above papers; while
the $G_1$ and $G_2$ terms were referred to as the ``uncorrelated variations''
in FWHM and luminosity, and they contribute to the virial uncertainty in
quadratic sum. The approach in \citet{Kelly_etal_2009a,Kelly_etal_2010}
implicitly assumed $\sigma_l^\prime=0$, while the approach in
\citet{Shen_etal_2008b} and \citet{Shen_Kelly_2010} is to set $\sigma_{\rm
corr}=0$ and consider non-zero $\sigma_l^\prime$. The latter choice is
motivated by the fact that the observed distribution of FWHM for SDSS quasar
samples is already narrow (dispersion $\lesssim 0.15$ dex) and the premise
that the virial uncertainty $\sigma_{\rm vir}$ should be no less than
$\sim0.3$ dex.

Physically $\sigma_l^\prime$ is unlikely to be zero. If this were true, it
would imply that single-epoch luminosity is an unbiased indicator for the
instantaneous BLR radius at fixed BH mass. While in practice it is more
natural to expect there are uncorrelated random scatter in both $L$ and $R$,
indicating a stochastic term in addition to the deterministic term (when
predicting $R$ with $L$), which will lead to biased estimates for $R$ at
fixed $L$. The sources of this stochastic term may include: a) the continuum
luminosity variation and response of the BLR is not synchronized; b)
individual quasars have different BLR properties; c) optical-UV continuum
luminosity is not as tightly connected to the BLR as the ionizing luminosity.
Furthermore, even if single-epoch luminosity were an unbiased indicator of
the instantaneous BLR radius, certain line width indicators (such as FWHM)
might still not response to all the variations in luminosity. For example,
consider a single BH where its luminosity varies (and its BLR radius varies
instantaneously following a perfect $R-L$ relation), and suppose that the
broad line is composed of a non-virialized component and a virialized
component. When luminosity increases (BLR expands), the virialized component
reduces line width, but the non-virialized component may increase line width
if it is originated from a radiatively driven wind (in the case of \CIV, more
blueshifted \CIV\ tends to have a larger FWHM, e.g., Shen et~al.\ 2008): the
combined line FWHM may not change due to the two opposite effects. Therefore
in this case although luminosity is tracing the BLR size perfectly, some
variation in luminosity is not compensated by variations in FWHM and should
be counted as the uncorrelated variation $\sigma_l^\prime$.

A non-zero $\sigma_l^\prime$ implies that the distribution of virial mass
estimates at fixed true mass {\em and} fixed luminosity, $p(m_e|m,l)$, is
different from the distribution of virial mass estimates at fixed true mass,
$p(m_e|m)$. In the extreme case where $\sigma_{\rm corr}=0$, i.e., FWHM does
not change in response to variations in luminosity at all, we have
\begin{equation}
m_e|m,l=m+0.5(l-\langle l\rangle_m)+2G_2(0|\sigma_{w})\ .
\end{equation}
Hence not only is the distribution $p(m_e|m,l)$ narrower than $p(m_e|m)$, but
also the expectation value of $m_e$ is biased from the true BH mass for any
fixed luminosities except for $l=\langle l\rangle_m$.

Now consider a more general form of the luminosity distribution at fixed true
mass and the actual slope in the observed luminosity-radius relation, we can
parameterize the distribution of $p(m_e|m,l)$ as:
\begin{equation}\label{eqn:B_corr}
m_e|m,l=m+\beta(l - \langle l\rangle_m)+\epsilon_{ml}\ ,
\end{equation}
where again $\langle l\rangle_m$ is the expectation value of luminosity at
fixed true mass, $\epsilon_{ml}$ is a random deviate with zero mean and
dispersion $\sigma_{ml}$, denoting the scatter of virial mass estimates {\em
at fixed true mass and fixed luminosity}, and the error slope $\beta$
describes the level of luminosity-dependent mass bias at fixed true mass and
luminosity. Both $\beta$ and $\epsilon_{ml}$ are to be constrained by our
data. Eqn.\ (\ref{eqn:B_corr}) implies that the variance of mass estimates at
fixed true mass and luminosity is reduced to:
\begin{equation}
{\rm Var}(m_e|m,l)={\rm Var}(m_e|m)(1-\rho^2)\ ,
\end{equation}
where $\rho^2=\beta^2{\rm Var}(l|m)/{\rm Var}(m_e|m)$, and ${\rm Var}(...)$
refers to the variance of a distribution. The formal uncertainty of the
virial mass estimator is then
\begin{eqnarray}
\sigma_{\rm vir}\equiv \sqrt{{\rm Var}(m_e|m)}&=&\sqrt{{\rm Var}(m_e|m,l) + \beta^2{\rm Var}(l|m)}\ .
\end{eqnarray}
If we assume a single log-normal distribution for $p(l|m)$ and
$\epsilon_{ml}$ (with a dispersion $\sigma_{ml}$), the above equation reduces
to
\begin{eqnarray}
\sigma_{\rm vir}=\sqrt{\sigma_{ml}^2+\beta^2\sigma_l^2}\ .
\end{eqnarray}
Note that here $\sigma_l$ is the total dispersion in $\log L$ at fixed mass,
rather than the portion $\sigma_l^\prime$ that is not responded by FWHM as in
Eqns.\ (\ref{eqn:dist_L}) and (\ref{eqn:p_me_m}).

Eqn.\ (\ref{eqn:B_corr}) is a rather generic form that describes the relation
between virial masses and true masses and the possible luminosity-dependent
bias in virial masses\footnote{One can work out a similar equation for the
distribution of $m_e$ at fixed $m$ and FWHM $w$,
$p(m_e|m,w)=m+\beta^\prime(w-\langle w\rangle_m)+\epsilon_{mw}$. A non-zero
$\sigma_w$ in Eqn.\ (\ref{eqn:dist_FWHM}) will lead to a non-zero
$\beta^\prime$ and $p(m_e|m,w)\neq p(m_e|m)$. However, this is of little
practical value since virial masses are never binned in FWHM.}, and is one of
the basic equations in our Bayesian approach. The value of $\beta$ depends on
the relative contributions from $\sigma_l^\prime$ and $\sigma_{\rm corr}$ in
the luminosity dispersion at fixed mass. Under the assumption that the {\em
mean} $R-L$ relation and a linear {\em mean} relation between FWHM and virial
velocity are correct as in the adopted virial estimators, a non-zero
$\sigma_l^\prime$ leads to a positive $\beta$. If the value of $\beta$
approaches the slope in the adopted mean $R-L$ relation, then it suggests
either luminosity or FWHM is a poor indicator for BLR size or virial velocity
over the narrow dynamical range at fixed true BH mass (although they could
still be reasonable indicators for large dynamical ranges in mass and
luminosity). On the other hand, if $\beta$ is small, then it means luminosity
and FWHM vary in concordance even over the narrow dynamical range at fixed
true BH mass, and hence are good indicators for BLR radius and virial
velocity. $\beta=0$ represents the extreme situation where FWHM responds to
all the variation in luminosity at fixed true mass (plus additional scatter
in FWHM), and no bias in virial masses is incurred when luminosity deviates
from $\langle l\rangle_m$. $\beta=0$ is generally assumed in most studies
with virial BH masses.

We also note that one advantage of using Eqn.\ (\ref{eqn:B_corr}) is that it
does not rely on the assumption that the mean $R-L$ relation and the linear
mean relation between FWHM and virial velocity used in these virial
estimators are correct. In other words, if the virial estimators adopted in
this work used incorrect forms for the mean $R-L$ relation and the mean
relation between FWHM and virial velocity, then a negative value of $\beta$
may be needed to correct $m_e$ at fixed $m$ and $l$. Of particular interest
here is whether or not radiation pressure is important in the dynamics of the
BLR \citep[e.g.,][]{Marconi_etal_2008}, which will indicate a negative
$\beta$ based on the virial masses that have not been corrected for radiation
pressure. We will test if a negative $\beta$ is required to model the
observed distribution in our Bayesian approach.

{\em Is there any indication for a non-zero $\beta$ from the reverberation
mapping AGN sample?} There are only $\sim 3$ dozens of RM AGNs and we do not
have enough objects with the same BH mass to test source-by-source
variations. Nevertheless, we can still test the luminosity-dependent bias
using repeated spectra for the same object when its luminosity changes a
significant amount over time. NGC 5548 is the most frequently monitored RM
AGN (\hbeta\ only), and has been observed in different luminosity states with
a spread of $\sim 0.5$ dex in luminosity
\citep[e.g.,][]{Peterson_etal_2004,Bentz_etal_2009b}, thus provides an ideal
test case for single-source variations.

\begin{figure*}
  \centering
    \includegraphics[width=0.45\textwidth]{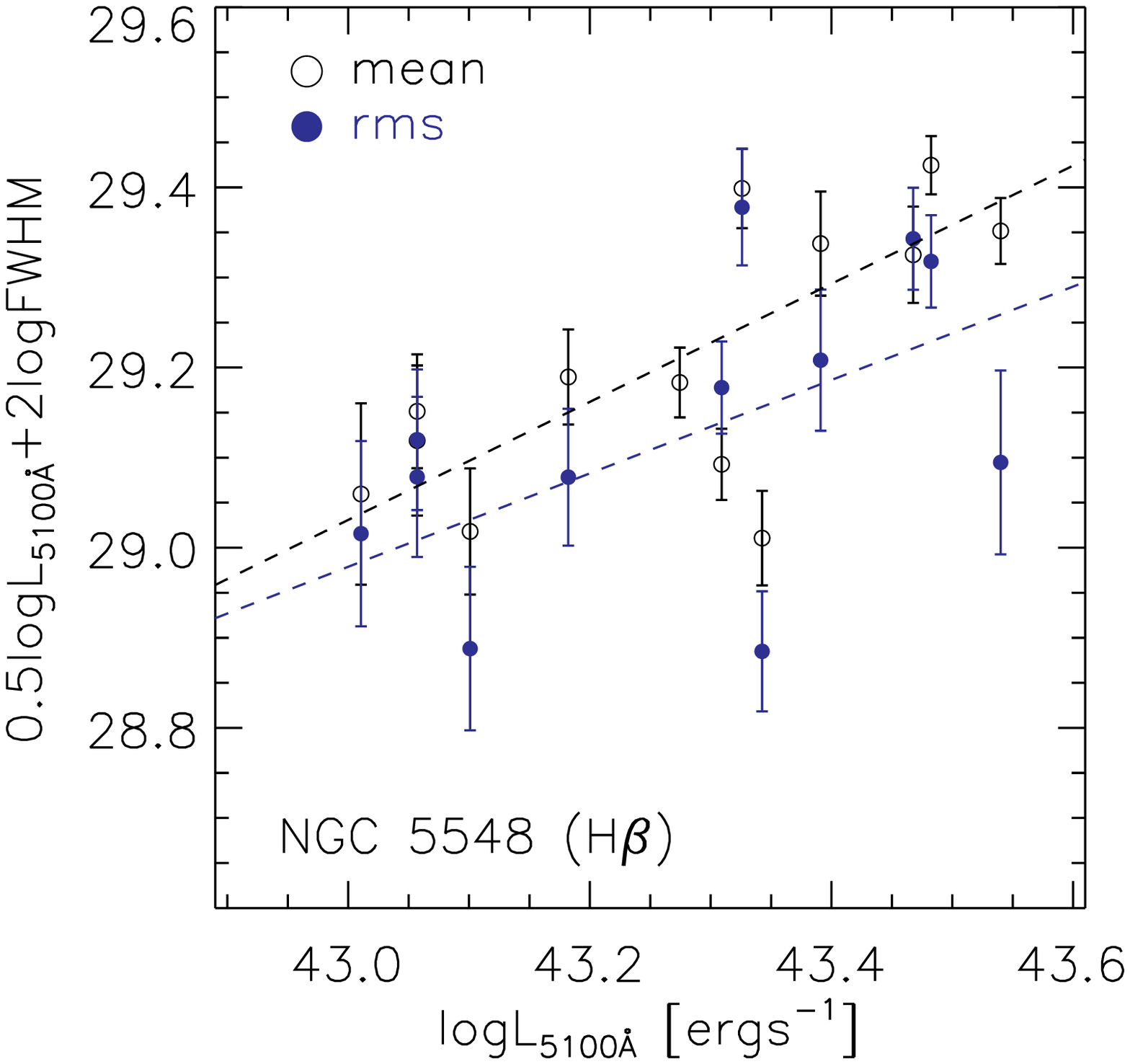}
    \includegraphics[width=0.45\textwidth]{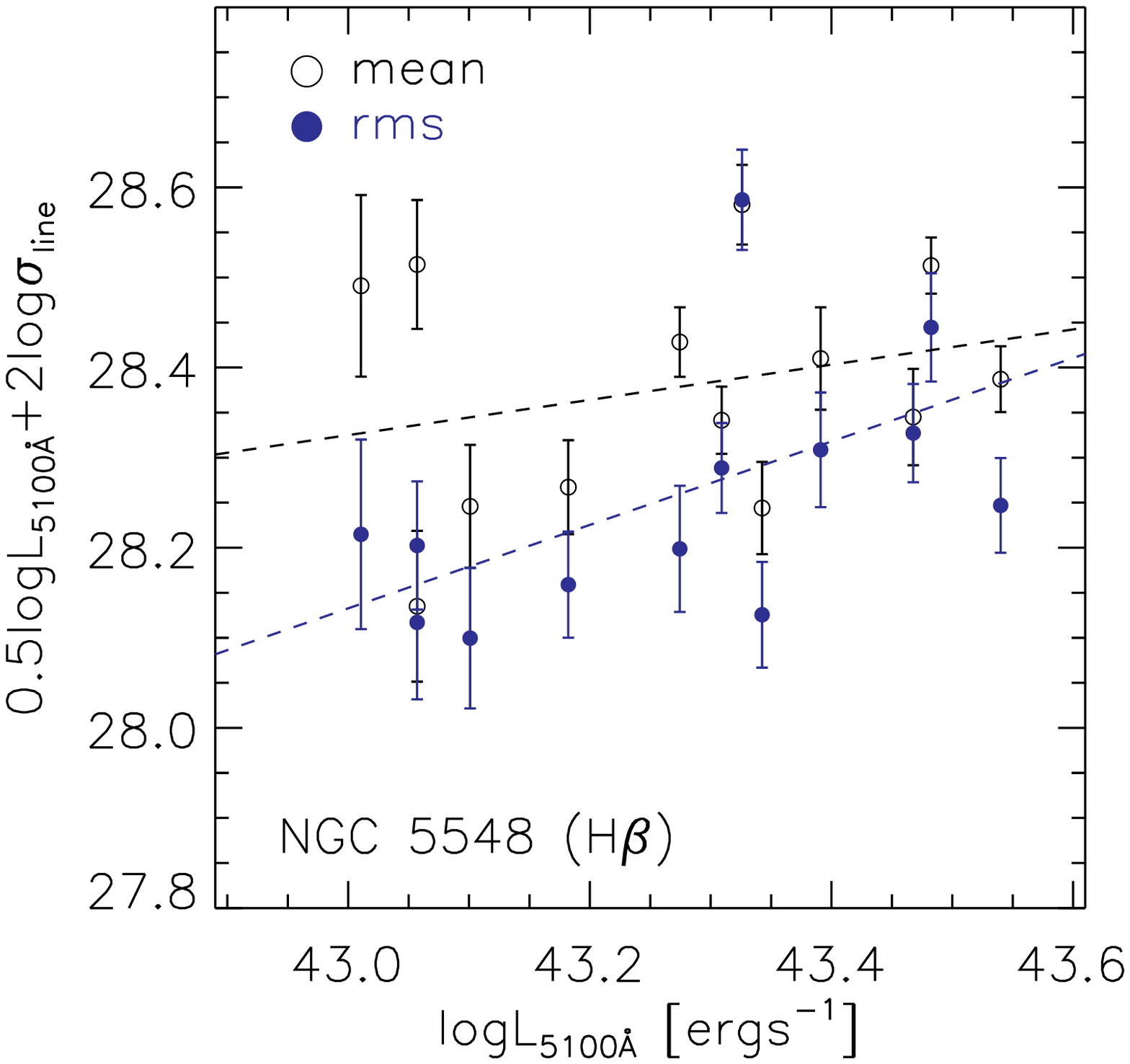}
    \caption{The dependence of the virial product computed from luminosity and line width as a function
    of luminosity, for a single object NGC 5548 and for \hbeta\ only. The data are from
    \citet{Collin_etal_2006}, and are based on both mean and rms spectra during each monitoring
    period. Error bars represent measurement errors. The error bars in luminosity have been
    omitted in the plot for clarity. The continuum luminosity has been corrected for
    host starlight using the correction provided by \citet{Bentz_etal_2009}. the black
    and blue dashed lines are the best linear-regression fits using the Bayesian method of \citet{Kelly_2007},
    for measurements based on mean and rms spectra, respectively. {\em Left}: virial product
    based on FWHM; the data point for Year 5 (JD 48954-49255) based on the rms spectrum has been
    suppressed due to problematic measurements \citep[e.g.,][]{Peterson_etal_2004,Collin_etal_2006}.
    {\em Right}: virial product based on line dispersion $\sigma_{\rm line}$.}
    \label{fig:ngc5548}
\end{figure*}

In Fig.\ \ref{fig:ngc5548} we show the \hbeta\ virial product for NGC 5548,
computed using the continuum luminosity and line width measured at different
luminosity states in each monitoring period, as a function of continuum
luminosity. The spectral measurements were taken from
\citet{Collin_etal_2006}, and we have corrected the continuum luminosity for
host starlight using the correction provided by \citet{Bentz_etal_2009}. The
line widths were measured from both the mean and rms
spectra\footnote{Strictly speaking, for single-epoch virial mass estimates,
neither the mean nor rms spectra are available. However, the spectral
variability during each monitoring period is small enough such that the mean
spectrum is close to single-epoch spectra within this period. } for each
monitoring period. The left and right panels of Fig.\ \ref{fig:ngc5548} show
the virial product computed using FWHM and line dispersion, respectively, and
its scaling with luminosity is the same as in the virial mass estimators
provided by \citet{Vestergaard_Peterson_2006}. The FWHM-based virial product
shows an average trend of increasing with luminosity, which means that FWHM
does not fully response to the variations in luminosity, leading to a
positive bias in the virial product (and thus in the virial mass estimate).
This trend seems to be slightly weaker when using line dispersion instead. A
linear regression analysis using the Bayesian method of \citet{Kelly_2007}
yields: $\beta\sim 0.65\pm0.27$ (FWHM, mean); $\beta\sim 0.51 \pm0.34$ (FWHM,
rms); $\beta\sim 0.20\pm0.30$ ($\sigma_{\rm line}$, mean); $\beta\sim 0.45
\pm 0.29$ ($\sigma_{\rm line}$, rms), where uncertainties are $1\sigma$.
While it is inconclusive based on this single object, there is some
indication that a positive $\beta$ is favored, especially for the virial
product based on FWHM from the mean spectra, which is the closest to that
based on FWHM from single-epoch spectra. It would be important to test this
for more objects with repeated spectra, and for \MgII\ and \CIV\ as well.

To summarize, because luminosity is an explicit term in virial mass
estimators, these virial mass estimates are no longer independent (and
unbiased) estimates of true masses when restricted to a narrow luminosity
range or a flux-limited sample, for cases where $\beta\neq 0$. Our view of
the uncertainties in these virial mass estimates (i.e., the scatter in
$p(m_e|m)$, as determined in the calibrations against the RM AGNs) is thus
different from that in \citet{Kollmeier_etal_2006} and
\citet{Steinhardt_Elvis_2010b}.

\subsubsection{Implementing the Bayesian Framework}

\begin{figure}
  \centering
    \includegraphics[height=0.5\textwidth,angle=90]{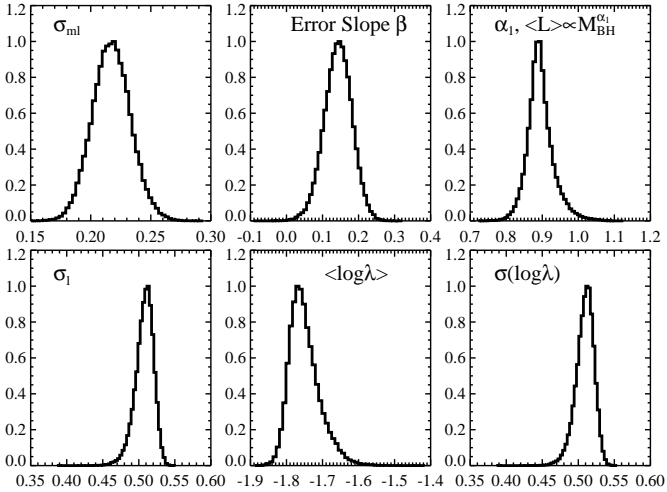}
    \caption{Model parameters for \texttt{zbin2}. Shown here are the posterior
    distributions of some model parameters and derived quantities. From top-left
    in clockwise order: the dispersion in mass estimates at fixed true mass
    {\em and} luminosity, $\sigma_{ml}$; the error slope $\beta$; the slope in
    the mean (true) mass-luminosity relation for our Eddington ratio model,
    $\alpha_1$; the dispersion in Eddington ratios for all broad-line
    quasars, $\sigma(\log\lambda)$ where $\lambda\equiv L_{\rm bol}/L_{\rm Edd}$;
    the mean Eddington ratio for all broad-line quasars, $\left<\log\lambda\right>$; the
    dispersion in luminosity at fixed true mass in our Eddington ratio model, $\sigma_l$.}
    \label{fig:zbin2_para_post}
\end{figure}

\begin{figure}
  \centering
    \includegraphics[height=0.5\textwidth,angle=90]{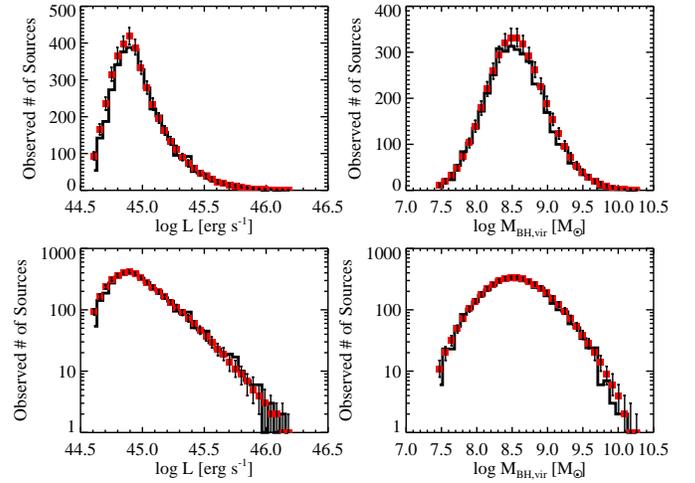}
    \caption{Posterior checks for \texttt{zbin2}. The solid black histograms
    shows the observed distributions. The red points and error bars are
    median results and uncertainties from our simulated samples using 500 random
    draws from the posterior distributions. The top and bottom panels show the
    histograms in linear and logarithmic scales, respectively. }
    \label{fig:zbin2_post_check}
\end{figure}

Now we proceed to describe our model setup and the implementation of the
Bayesian framework. Below we describe the basics of our model approach. More
details regarding the Bayesian approach can be found in
\citet{Kelly_etal_2009a} and \citet{Kelly_etal_2010}.

\begin{enumerate}

\item[a.] {\em The BHMF and luminosity distribution model}. As in
    \citet{Kelly_etal_2009a,Kelly_etal_2010}, we use a mixture of
    log-normal distributions as our model for the true BHMF, and a single
    log-normal luminosity (Eddington ratio) distribution at fixed true BH
    mass. The mixture of log-normals is flexible enough to capture the
    basic shape of any physical BHMF, and greatly simplifies the
    computation as many integrations can be done analytically. The model
    true BHMF reads
    \begin{eqnarray}\label{eqn:bhmf}
    \Phi(m)=N\left(\frac{dV}{dz}\right)^{-1}\sum_{k=1}^K\frac{\pi_k}{\sqrt{2\pi \sigma_k^2}}\exp\left[-\frac{(m-\mu_k)^2}{2\sigma_k^2}\right]\ ,
    \end{eqnarray}
    where $m\equiv \log M_{\rm BH}$, $N$ is the total number of quasars,
    $\mu_k$ and $\sigma_k$ are the mean and dispersion of the $k$th
    Gaussian, and $\sum_{k=1}^K\pi_k=1$. We use $K=3$ log-normals to
    describe the BHMF, as we do not find significant difference when
    increasing the number of log-normals used. The luminosity
    distribution at fixed BH mass is modeled as
    \begin{equation}\label{eqn:edding_model}
    p(l|m)=\frac{1}{\sqrt{2\pi\sigma_l^2}}\exp\left(-\frac{[l-\alpha_0-\alpha_1(m-9)]^2}{2\sigma_l^2}\right)\ ,
    \end{equation}
    where $l\equiv \log L$, $\alpha_0$ and $\alpha_1$ describe a
    mass-dependent mean luminosity, and $\sigma_l$ is the scatter in
    luminosity at fixed mass. The LF is therefore
    \begin{equation}\label{eqn:lf}
    \Phi(l)=\int \Phi(m)p(l|m)dm\ .
    \end{equation}

\item[b.] {\em The virial mass prescription}. We assume that virial
    masses are unbiased when averaged over luminosity at fixed true mass
    (i.e., Eqns.\ \ref{eqn:p_me_m},\ref{eqn:B_corr}), and we generate
    virial masses at fixed true mass and luminosity according to Eqn.\
    (\ref{eqn:B_corr}), assuming a single Gaussian (with dispersion
    $\sigma_{ml}$) to describe the scatter $\epsilon_{ml}$ at fixed mass
    and luminosity.

\item[c.] {\em The redshift distribution}. Because the redshift bins are
    narrow, we approximate the distribution of redshifts across the bin
    as a power-law, where the power-law index $\gamma$ is a free
    parameter:
  \begin{equation} \label{eqn:zdist}
    p(z|\gamma) = \frac{(1+\gamma)z^{\gamma}}{z_{\rm max}^{1+\gamma} - z_{\rm min}^{1+\gamma}}.
\end{equation}
Here, $z_{\rm max}$ and $z_{\rm min}$ define the upper and lower boundary
of the redshift bin, respectively.

\item[d.] {\em The posterior distribution} $p(\theta|m_e, l,z)$ is
    \begin{equation}\label{eqn:post}
    p(\theta|m_e,l,z) \propto p(\theta)[p(I=1|\theta)]^{-N}\prod_{i=1}^N p(m_{e,i},l_i,z_i|\theta) \\
  \end{equation}
    where $\theta(\pi_k,\mu_k,
    \sigma_k,\alpha_0,\alpha_1,\sigma_l,\beta,\sigma_{ml},\gamma)$ is the
    set of model parameters, $N$ is the total number of quasars,
    $p(\theta)$ is the prior on $\theta$, $p(I=1|\theta)$ is the
    probability as a function of $\theta$ that a broad-line quasar is
    included in the flux-limited SDSS quasar sample, and the likelihood
    function $p(m_{e,i},l_i,z_i|\theta)$ is determined by Eqns.
    (\ref{eqn:B_corr}), (\ref{eqn:bhmf}), (\ref{eqn:edding_model}) and
    (\ref{eqn:zdist}).

\end{enumerate}

In this work, we derive the continuum luminosity at $2500$\,\AA, $l\equiv
\log L$, from the $i$-band magnitude according to the prescription given in
\citet{Richards_etal_2006a}. This is a departure from the approach taken by
\citet{Kelly_etal_2010}, who used the continuum luminosity estimated by
\citet{Vestergaard_etal_2008}. However, as explained in
\citet{Kelly_etal_2010}, because the SDSS selection function is in terms of
the $i$-band magnitude, they had to assume a model for the distribution of
$i$ at fixed luminosity, and then calculate the selection function in terms
of luminosity by averaging over this model distribution. They note that this
approach can lead to instability in the estimated mass function in mass bins
that are severely incomplete, as small errors in the selection function can
lead to large deviations in $p(I=1|\theta)$, which appears in the denominator
in Equation (\ref{eqn:post}). Instead, we use the luminosity derived from the
$i$-band magnitude to ameliorate this effect, as the selection function is
calculated in terms of $i$.

It is necessary to impose some prior constraints on $\beta$ and $\sigma_{ml}$
based on the reverberation mapping data set, as these parameters are
degenerate with some of the other parameters. Unlike most previous work, we
do not fix the values of $\beta$ and $\sigma_{ml}$ to, say, $\beta = 0$ and
$\sigma_{ml} = 0.4$ dex, but use a prior distribution which incorporates our
uncertainty in these parameters. This uncertainty will be reflected in the
probability distribution of the mass function, given the SDSS DR7 data set.
We applied the Bayesian linear regression method of \citet{Kelly_2007} to the
reverberation mapping sample in order to estimate the probability
distribution of $\beta$ and $\sigma_{ml}$ based on this sample. The method of
\citet{Kelly_2007} incorporates the measurement errors in the mass estimates,
which is important when estimating the amplitude of the scatter in the mass
estimates. We set $\langle l \rangle_m$ equal to the mean luminosity for the
reverberation mapping sample\footnote{Ideally one should use a population of
objects with the same true BH mass to perform linear regression using Eqn.\
(\ref{eqn:B_corr}), which is not available given the limited size of the RM
sample. So instead we use the whole RM sample. Nevertheless, the derived
prior constraint on $\beta$ is weak and our results are insensitive on the
prior on $\beta$. }. We used the values of RM black hole mass (assumed to be
true masses) given by \citet{Peterson_etal_2004}. For the H$\beta$
calibration, we used the value of $5100$\,\AA\ luminosity given in
\citet{Bentz_etal_2009} and value of FWHM given in
\citet{Vestergaard_Peterson_2006}; when there were multiple measurements, we
averaged them together. For \CIV\ we used the values given in
\citet{Vestergaard_Peterson_2006}. For H$\beta$ we found that $\beta = 0.16
\pm 0.1$, and that the posterior distribution for $\sigma^2_{ml}$ is well
described by a scaled inverse $\chi^2$ distribution with $\nu \approx 20$
degrees of freedom and scale parameter $s^2 = 0.1$. For \CIV\ we found that
$\beta = 0.15 \pm 0.14, \nu \approx 20, s^2 = 0.17$. For \CIV\ this is
similar to the usually quoted scatter in the mass estimates of $s \approx
0.4$ dex, but the scatter is less for H$\beta$, $s \approx 0.3$ dex. This
reduced uncertainty is likely because we have used 5100\,\AA\ luminosity
values which are corrected for host galaxy starlight \citep{Bentz_etal_2009}.

\begin{figure}
  \centering
    \includegraphics[height=0.5\textwidth,angle=90]{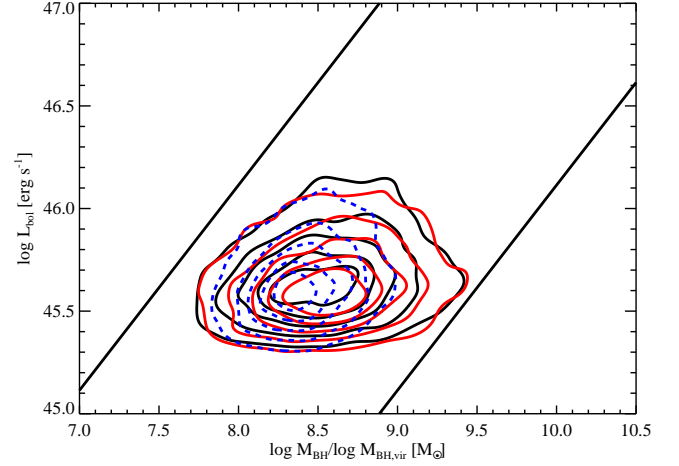}
    \caption{Posterior checks for \texttt{zbin2} in the mass-luminosity plane above
    the flux limit. The two black lines indicate Eddington ratios of $0.01$ and $1$.
    The black and red contours are for the observed and simulated distributions
    using virial masses, and the blue dashed contour shows the simulated distribution
    with true masses. Our model fits the observed distribution well, and the
    distribution using true masses is different from that using virial masses.}
    \label{fig:zbin2_ml_plane}
\end{figure}

\begin{figure}
  \centering
    \includegraphics[height=0.5\textwidth,angle=90]{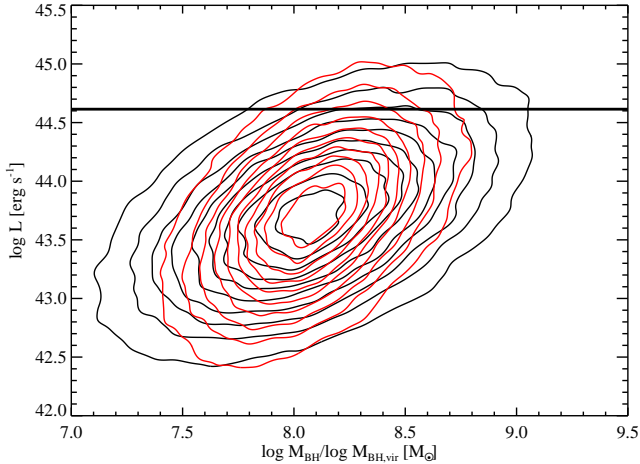}
    \caption{The simulated mass-luminosity plane for \texttt{zbin2}, which extends
    below the flux limit (the black horizontal line). The red contour is the
    distribution based on true BH masses, and is determined by our model BHMF and
    Eddington ratio model. The black contour is the distribution based on
    virial BH masses. The flux limit only selects the most luminous object into
    our sample, and the distribution based on virial BH masses is flatter than the
    one based on true masses due both to the scatter $\sigma_{ml}$ and a non-zero
    $\beta$ (see Eqn.\ \ref{eqn:B_corr}). }
    \label{fig:zbin2_ml_plane_all}
\end{figure}

Based on the reverberation mapping results, we impose a Cauchy prior
distribution on $\beta$ with mean and scale parameters equal to those derived
from the reverberation mapping data, and we impose a scaled-inverse $\chi^2$
prior distribution on $\sigma^2_{ml}$ with $\nu = 15$ degrees of freedom and
scale parameter set to that from the reverberation mapping sample. For \MgII\
we use the values derived for \hbeta\ since the single-epoch virial masses
based on the two lines seem to correlate with each other well
\citep[e.g.,][]{Shen_etal_2011a}. We have used a Cauchy prior because it has
significantly more probability in the tails than the usual Gaussian
distribution, making our prior assumptions more robust. Similarly, we reduced
the degrees of freedom for the prior on $\sigma^2_{ml}$ compared to the
reverberation mapping sample in order to increase our prior uncertainty on
$\sigma_{ml}$. This choice of prior assumes an uncertainty on $\sigma_{ml}$
of $\approx 20\%$.

As in \citet{Kelly_etal_2009a} and \citet{Kelly_etal_2010}, we use a Markov
Chain Monte Carlo (MCMC) sampler algorithm to obtain random draws of $\theta$
according to Eqn.\ (\ref{eqn:post}) and thus the posterior distribution of
model parameters given the observed data in the virial mass-luminosity plane.
Our MCMC sampler employs a combination of Metropolis-Hastings updates with
parallel tempering. The reader is referred to \citet{Kelly_etal_2009a} and
\citet{Kelly_etal_2010} for further details.

Different from \citet{Kelly_etal_2009a} and \citet{Kelly_etal_2010}, we model
the data in individual redshift bins instead of for the whole sample. We
treat each redshift bin as an independent data set. This allows us to explore
possible redshift evolution of our model parameters, as well as the
sensitivity on changes in the detection luminosity threshold and the specific
virial mass estimator used. The caveat is that the constraints are generally
weaker given less data points in each bin, and that the constrained
parameters do not necessarily vary smoothly across adjacent redshift bins.

\section{Results of the Bayesian Approach}\label{sec:results}

\subsection{\texttt{zbin2} as an example}\label{sec:zbin2}

We use \texttt{zbin2} as an example to demonstrate the information that we
can retrieve from the posterior distributions. This bin uses the most
reliable \hbeta\ line to estimate virial BH masses; it also has negligible
host galaxy contamination compared with \texttt{zbin1}. Therefore the
constraints for this bin are expected to be the most robust.

Fig.\ \ref{fig:zbin2_para_post} shows the posterior distributions of our
model parameters for \texttt{zbin2}. These parameters are tightly
constrained, although degeneracy does exist among these parameters.

\begin{figure}
  \centering
    \includegraphics[width=0.48\textwidth]{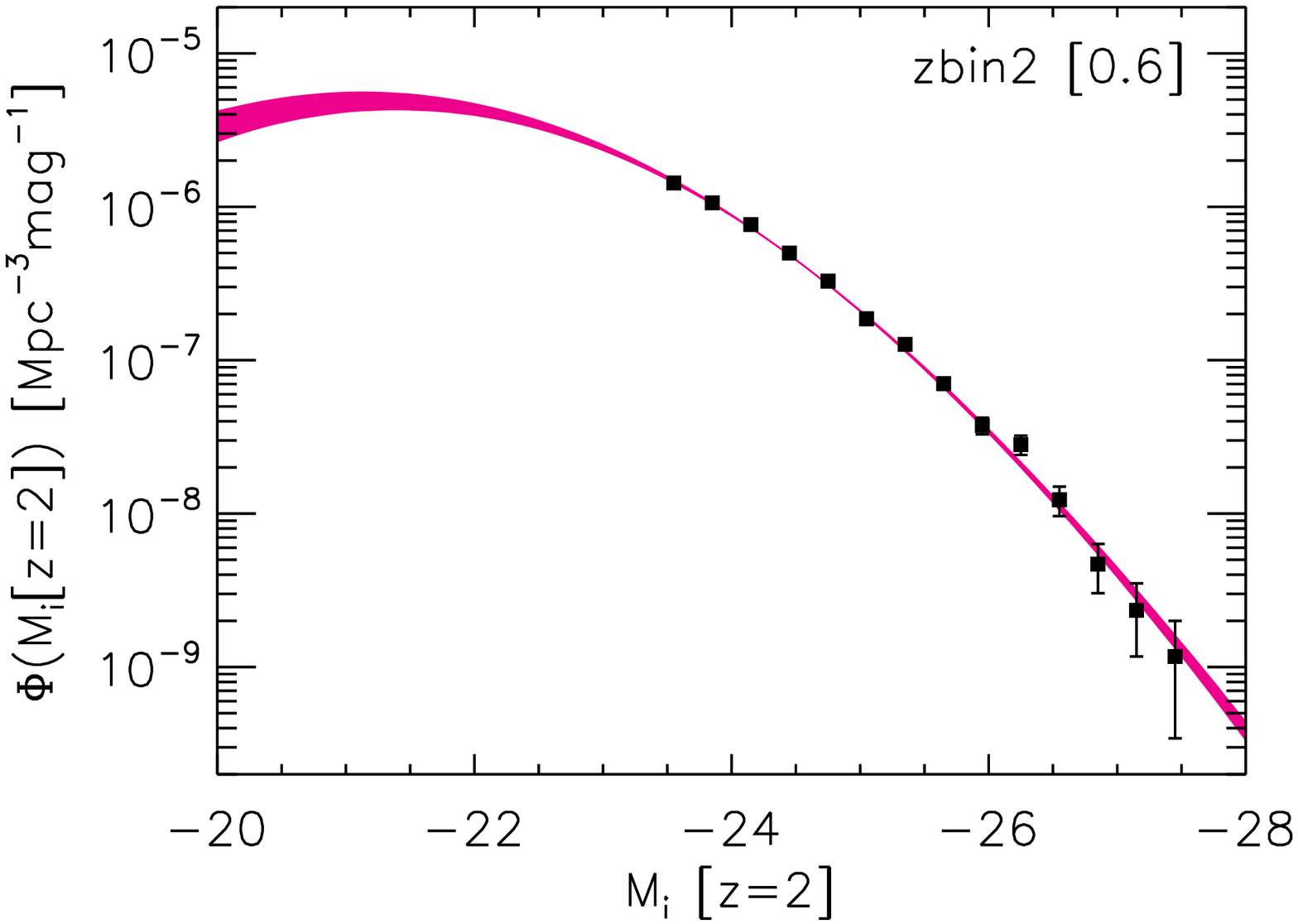}
    \includegraphics[width=0.48\textwidth]{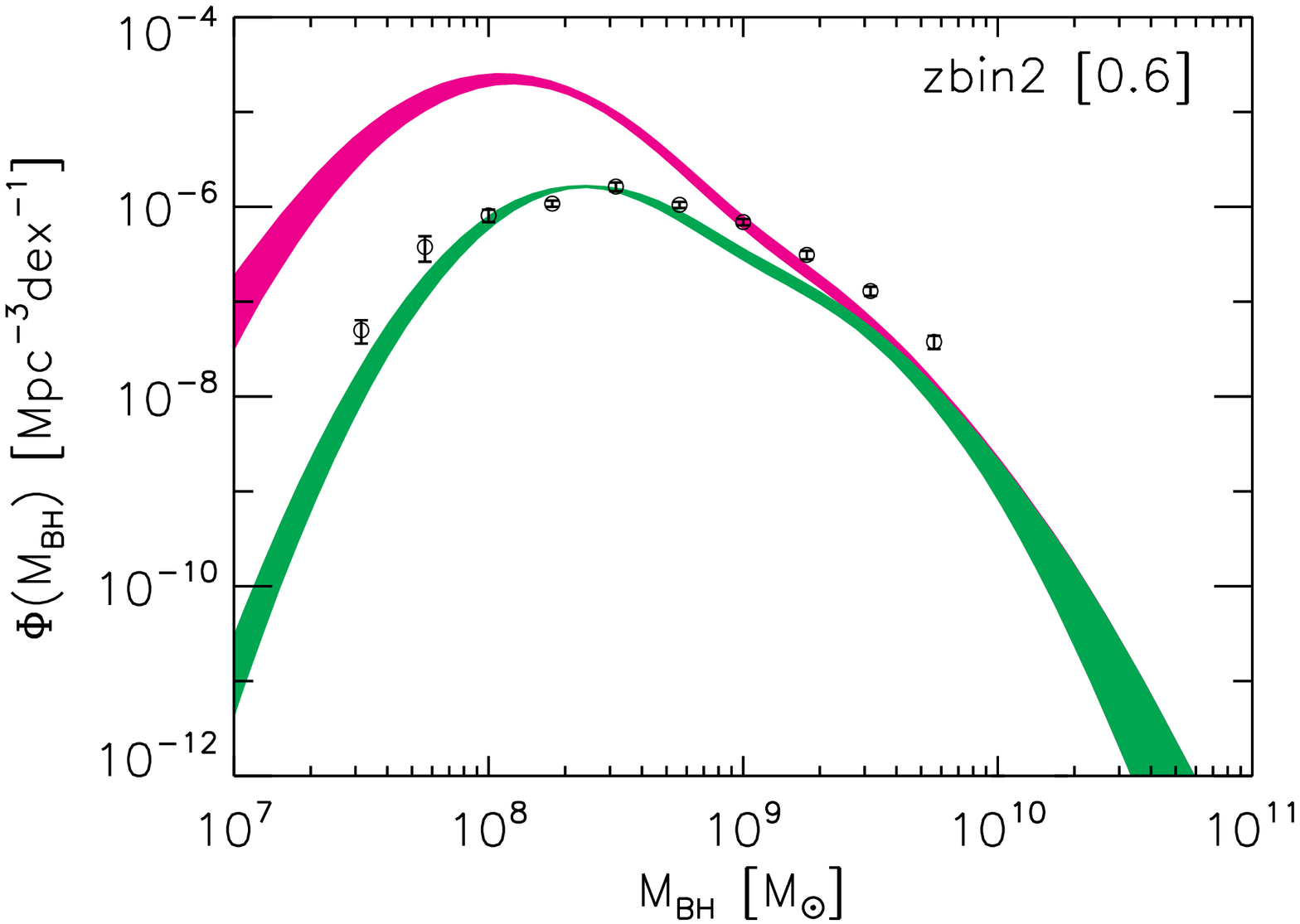}
    \caption{Model LF (top panel) and BHMF (bottom panel) for \texttt{zbin2}.
    The data points and error bars are the binned LF and virial BHMF estimated in
    \S\ref{subsec:bin_BHMF}. The color shaded regions are the 68\% percentile range from
    our model LF and BHMF. In the bottom panel, the green shaded region is for the
    detectable (i.e., above the flux limit) true BHMF, and the magenta one is for
    all the broad-line quasars. The turnover of the magenta line below
    $\sim 10^8\,M_\odot$ is a feature constrained by the data and our model, i.e.,
    if there were more lower mass BHs, it would be difficult to fit the observed distribution
    in the mass-luminosity plane (cf. Fig.\ \ref{fig:zbin2_ml_plane_all}).
    }
    \label{fig:zbin2_lf_bhmf}
\end{figure}

\begin{figure}
  \centering
    \includegraphics[width=0.48\textwidth]{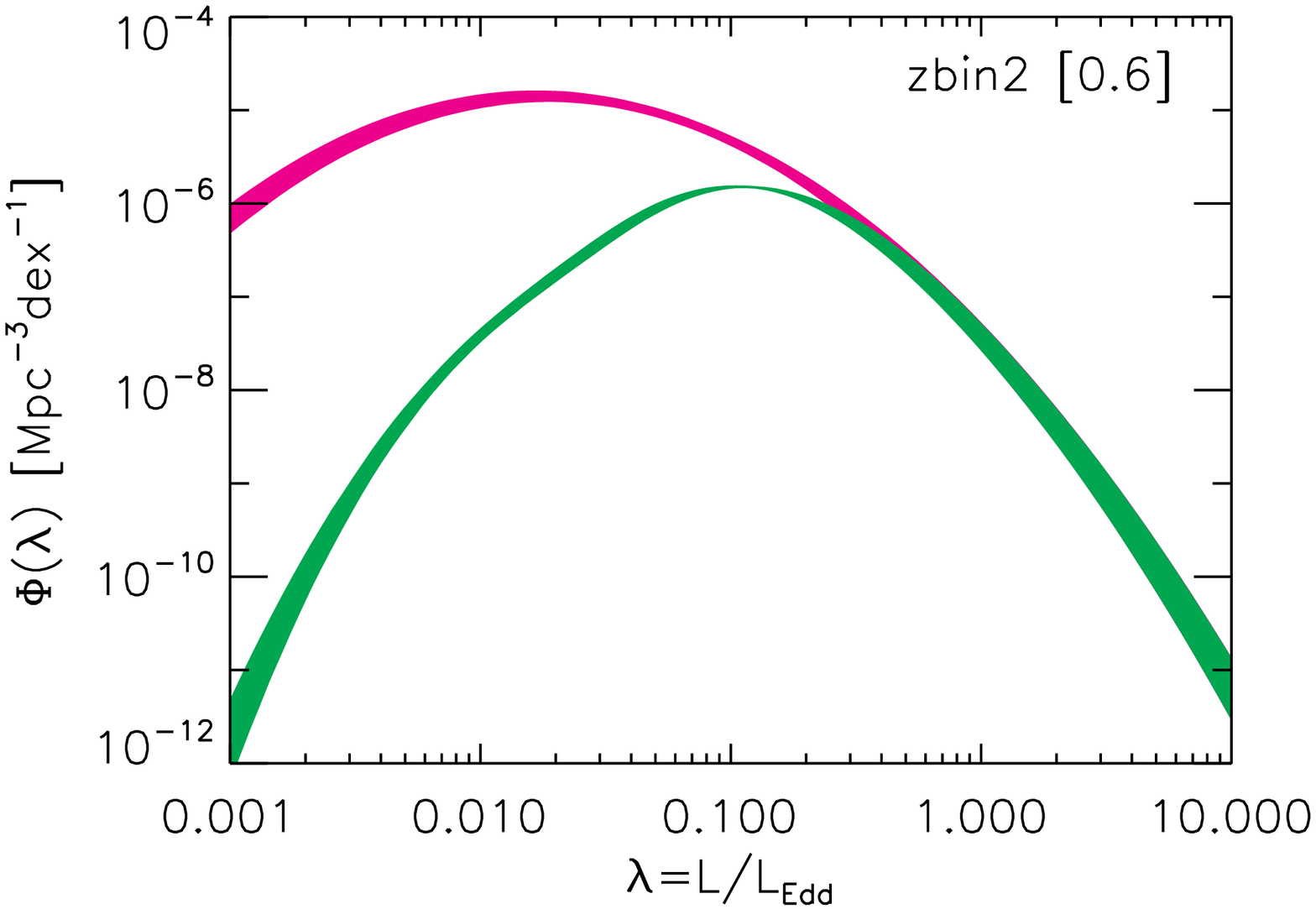}
    \includegraphics[width=0.48\textwidth]{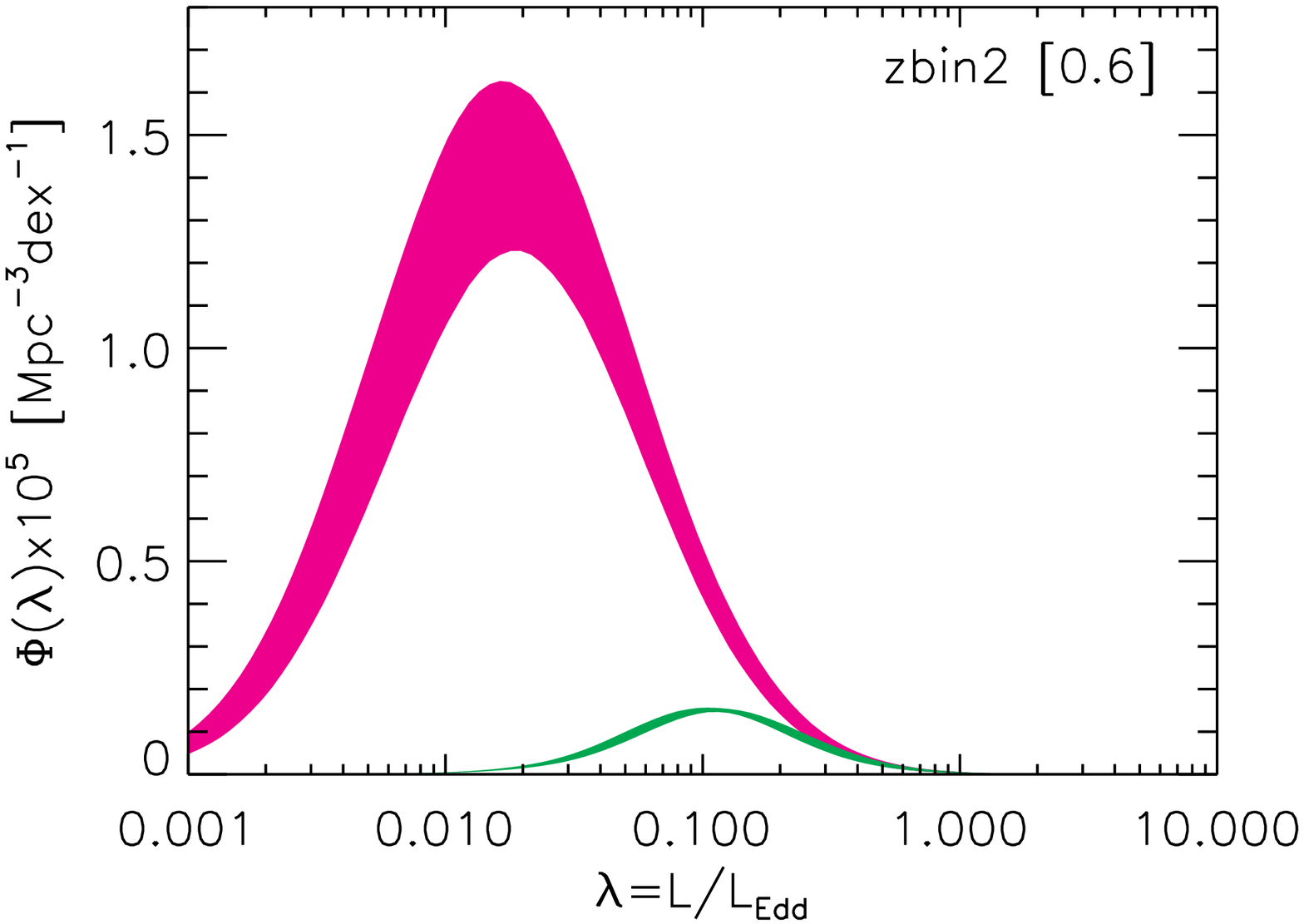}
    \caption{Eddington ratio function for \texttt{zbin2}. As in the bottom panel of
    Fig.\ \ref{fig:zbin2_lf_bhmf}, the green and magenta shaded regions are the
    68\% percentile range for the detectable and all broad-line quasars, calculated
    using our model posterior distributions. The top and bottom panels show the
    Eddington ratio function in logarithmic and linear scales, respectively. While
    most broad-line quasars have a mean Eddington ratio of $\sim 0.02$ for this redshift,
    the detected ones have a higher mean Eddington ratio of $\sim 0.1$ due to the
    Malmquist-type bias. }
    \label{fig:zbin2_edd}
\end{figure}

Fig.\ \ref{fig:zbin2_post_check} presents the posterior checks of our model
against the data. The black histograms show the distribution of observed
luminosities and virial masses. The points and error bars are the median and
68\% percentile for simulated samples generated using random draws from the
posterior distribution. This ensures that our model reproduces the observed
luminosity and virial mass distributions. Fig.\ \ref{fig:zbin2_ml_plane}
further shows the comparison between model prediction and data in the
two-dimensional mass-luminosity plane, where the model is the one that has
the maximum posterior probability. The black and red contours show the
observed and model-predicted joint-distributions of luminosity and virial
mass in our sample, while the blue contours show that for the true masses.
The true masses are scattered and biased according to Eqn.\
(\ref{eqn:B_corr}). For this particular bin the luminosity-dependent bias
between virial and true masses is only moderate ($\beta\sim 0.15$ as seen in
Fig.\ \ref{fig:zbin2_para_post}).

Fig.\ \ref{fig:zbin2_ml_plane_all} summarizes the relation between luminosity
and BH mass, the effects of the flux limit, and the difference between virial
masses and true masses in the mass-luminosity plane. Black and red contours
indicate the simulated distributions using virial masses and true masses,
respectively. Quasars in this \texttt{zbin} peak around $10^8\,M_\odot$ with
sub-Eddington ratios. This turn-over at low mass is a feature {\em as
constrained by our model}. If the low-mass end BHMF continued to rise, there
would be more smaller BHs with high Eddington ratio being scattered into our
sample, and it would be difficult to fit the observed distribution above the
detection threshold (the black horizontal line). However, since the low-mass
end is only constrained by a few objects close to the flux limit, the
location of the turn-over will most likely depend on our model assumptions,
such as the mixed-Gaussian model for the underlying BHMF and the simple
log-normal Eddington ratio distribution at fixed true mass; the assumption
that the error distribution of virial masses is Gaussian may also affect the
result. To fully tackle these issues, deeper data is needed to provide more
stringent constraints at the low-mass end, and different models (such as a
mixture of power-laws for the true BHMF), albeit more computationally
challenging, are required to test the robustness of this turn-over. Therefore
we do not claim a robust detection of a turn-over in the true BHMF in this
work, but simply note the possibility of such a turn-over, and point out that
the turn-over at larger masses seen in the {\em flux-limited} BHMF is a
selection effect (see below).

The luminosity at fixed BH mass has a substantial scatter. However, the SDSS
flux limit only picks up the luminous objects. Since there are more low-mass
BHs with high Eddington ratios being scattered into our sample, the Eddington
ratio distribution is subject to a Malmquist-type bias
\citep[e.g.,][]{Eddington_1913,Malmquist_1922,Lauer_etal_2007,Shen_etal_2008b,Kelly_etal_2009a,Kelly_etal_2010}.
Furthermore, these BH masses above the detection threshold are estimated as
virial BH masses according to Eqn.\ (\ref{eqn:B_corr}), which further
stretches the distribution horizontally in the mass-luminosity plane. The
flux limit, the scatter in $m_{e}$ at fixed $m$ ($\sigma_{ml}$), and the
luminosity-dependent bias (inferred by the value of $\beta$), have changed
the distribution in the mass-luminosity plane substantially. There is much
weaker evidence for a ``sub-Eddington boundary'' claimed by
\citet{Steinhardt_Elvis_2010a} if we use true masses instead of virial
masses, and there is no need to invoke alternative virial calibrations to
remove this boundary \citep{Rafiee_Hall_2011a}. However, such a boundary may
exist if the current versions of virial estimators are systematically biased,
or if our model parameterization is inappropriate. A more detailed
investigation using a different model parameterization of the mass-luminosity
plane will be presented elsewhere (Kelly \& Shen, in preparation; see
discussions in \S\ref{sec:disc_caveats}).

\begin{figure*}
  \centering
    \includegraphics[width=0.95\textwidth]{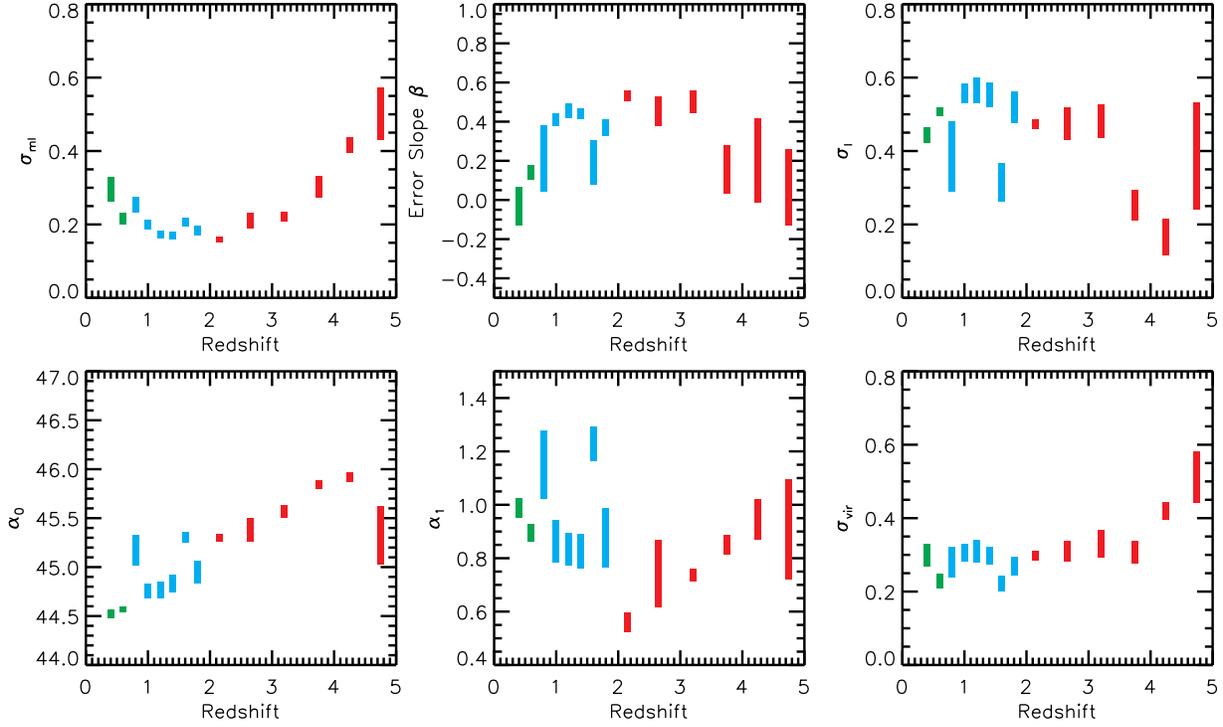}
    \caption{Model parameters for all 14 \texttt{zbins}. The green, cyan and red colors
    are for the \hbeta, \MgII\ and \CIV\ respectively. The vertical extent of each segment
    indicate the 68\% percentile range from our posterior distributions. }
    \label{fig:model_evo}
\end{figure*}

Fig.\ \ref{fig:zbin2_lf_bhmf} shows the model LF ({\em upper}) and BHMF ({\em
bottom}) and comparisons with data. The shaded curves indicate the 68\%
percentile range. The LF is well constrained even when extrapolated to
fainter luminosities than probed by the SDSS sample. The true BHMF is plotted
in magenta. It turns over below $\sim 10^8\,M_\odot$, as already noted in
Fig.\ \ref{fig:zbin2_ml_plane_all}. The green shaded curve indicate the {\em
detected} BHMF, i.e., the population of quasars that have instantaneous
luminosities above the flux limit of the SDSS sample. The flux limit causes
significant selection incompleteness in terms of BH mass, which becomes worse
towards smaller BHs. As a consequence, the turn-over in the {\em detected}
BHMF shifts to a larger BH mass. We also over-plotted the binned virial BHMF
for our flux-limited sample in open circles. The Poisson errors in the virial
BHMF are not meaningful, and significantly underestimate the uncertainty in
the BHMF. The shape of the virial BHMF also differs from the green curve, due
to the difference between true and virial BH masses\footnote{We note that in
general $dm_e\neq dm$, so the integrated area under the green curve and the
virial BHMF does not necessarily agree. But the total number of ``detected''
quasars in our model is the same as observed.}.

Fig.\ \ref{fig:zbin2_edd} shows the Eddington ratio function of all broad
line quasars in magenta, and of those detectable quasars in green. Most
quasars are accreting below Eddington, and the Eddington ratio distribution
for the detected quasars is biased high due to selection effect. There is a
significant dispersion ($\sim 0.5$ dex) in Eddington ratios at fixed BH mass
and for the entire quasar population.

\subsection{Results in all \texttt{zbins}}\label{sec:all_zbins}

Fig.\ \ref{fig:model_evo} shows the $68\%$ percentile range of parameters
$\sigma_{ml}$, $\beta$, $\sigma_l$, $\alpha_0$, $\alpha_1$, and $\sigma_{\rm
vir}=\sqrt{\sigma_{ml}^2+\beta^2\sigma_l^2}$. The green, cyan and red colors
indicate the \hbeta, \MgII\ and \CIV\ mass estimators, respectively. The
constraints on some parameters (such as $\beta$, $\sigma_l$ and $\alpha_1$)
are poorer than the others, and there is generally degeneracy among these
parameters. Also, due to the changes of detection luminosity threshold with
redshift and the complexity of the multi-dimensional parameter space,
adjacent redshift bins do not necessarily make smooth transition, even if we
stick to one mass estimator.

One purpose of our Bayesian approach is to explore if there is evidence for a
luminosity-dependent bias in virial BH masses through the parameter $\beta$
in Eqn.\ (\ref{eqn:B_corr}), to be constrained by the data. We found that for
\hbeta\ there is generally no need for such a bias. However, for \MgII\ and
\CIV, there is some evidence that the best models favor a positive $\beta$
value between $0.2$ and $0.6$. Because of this positive $\beta$ value, the
average virial BH masses are biased high by a factor of a few at $z\ga 0.7$.
Including this luminosity-dependent bias, the uncertainty in virial mass
estimates is $\sigma_{\rm vir}\ga 0.3$ dex and larger than the scatter
$\sigma_{ml}$ at fixed luminosity and true mass. We note that this positive
luminosity bias is in the opposite sense to the bias proposed by
\citet{Marconi_etal_2008}, who argue that these virial estimators do not
include the effect of radiation pressure and hence are biased low at high
luminosities. This would argue for a negative value of $\beta$, which is not
seen in our MCMC results.

The dispersion in luminosity at fixed BH mass is constrained to be $0.2\la
\sigma_l\la 0.6$ with a median value of $\sim 0.4$ dex. The slope in the mean
luminosity and mass relation is constrained to be $0.6\la \alpha_1\la 1.2$,
and there is evidence that the normalization $\alpha_0$ increases with
redshift. These constraints are weaker than earlier work
\citep{Kelly_etal_2009a,Kelly_etal_2010}, due to the additional freedom
introduced by the luminosity-dependent bias and the fact that we are now
modeling the data in individual redshift bins independently.

\begin{figure*}
  \centering
    \includegraphics[height=0.9\textwidth]{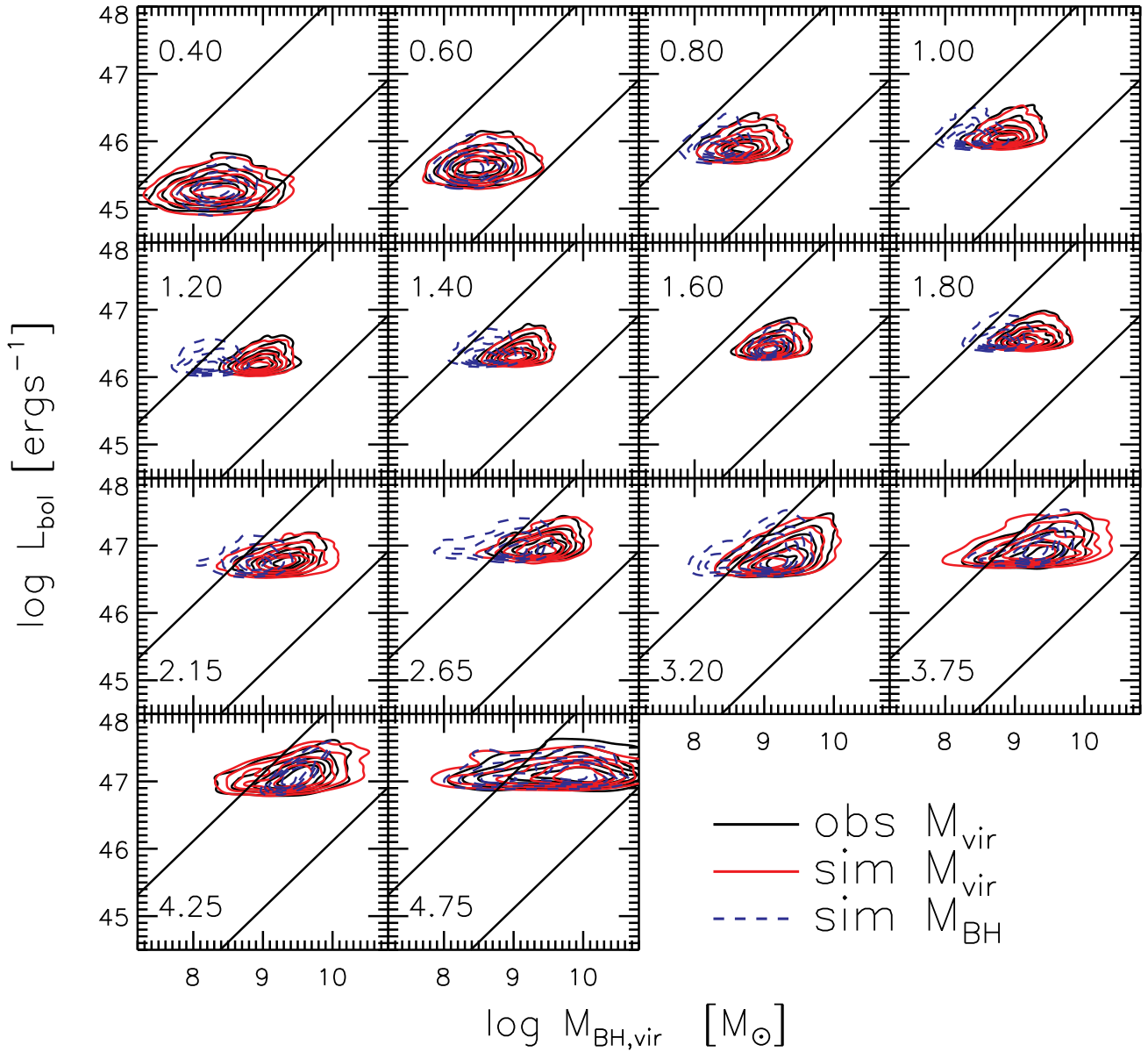}
    \caption{Posterior checks for all \texttt{zbins} in the mass-luminosity plane above the detection threshold. The black contours
    are for the observed quasars with virial masses; the red and blue contours are for the simulated quasars using
    the model realization with the maximum posterior probability, with virial masses and true masses, respectively.
    The two straight lines indicate Eddington ratios $\lambda=0.01$ and 1. The redshift is marked in each panel. Our
    model reproduces the observed distributions well, and demonstrates the difference between true masses and
    virial masses.
    }
    \label{fig:2d_ml_plane}
\end{figure*}

Finally, as a sanity check, Fig.\ \ref{fig:2d_ml_plane} shows the comparison
between model and data for all 14 \texttt{zbins} in the mass-luminosity
plane, where again the model is the one that has the maximum posterior
probability. In all \texttt{zbins} the observed distribution is reproduced.
The true mass distribution is generally different, but the level of this
difference varies from bin to bin. The latter reflects the large uncertainty
of determining the relation between true mass and virial mass (i.e., Eqn.\
\ref{eqn:B_corr}) when the luminosity threshold and line estimator change
with redshift.

\begin{figure*}
  \centering
    \includegraphics[width=0.9\textwidth]{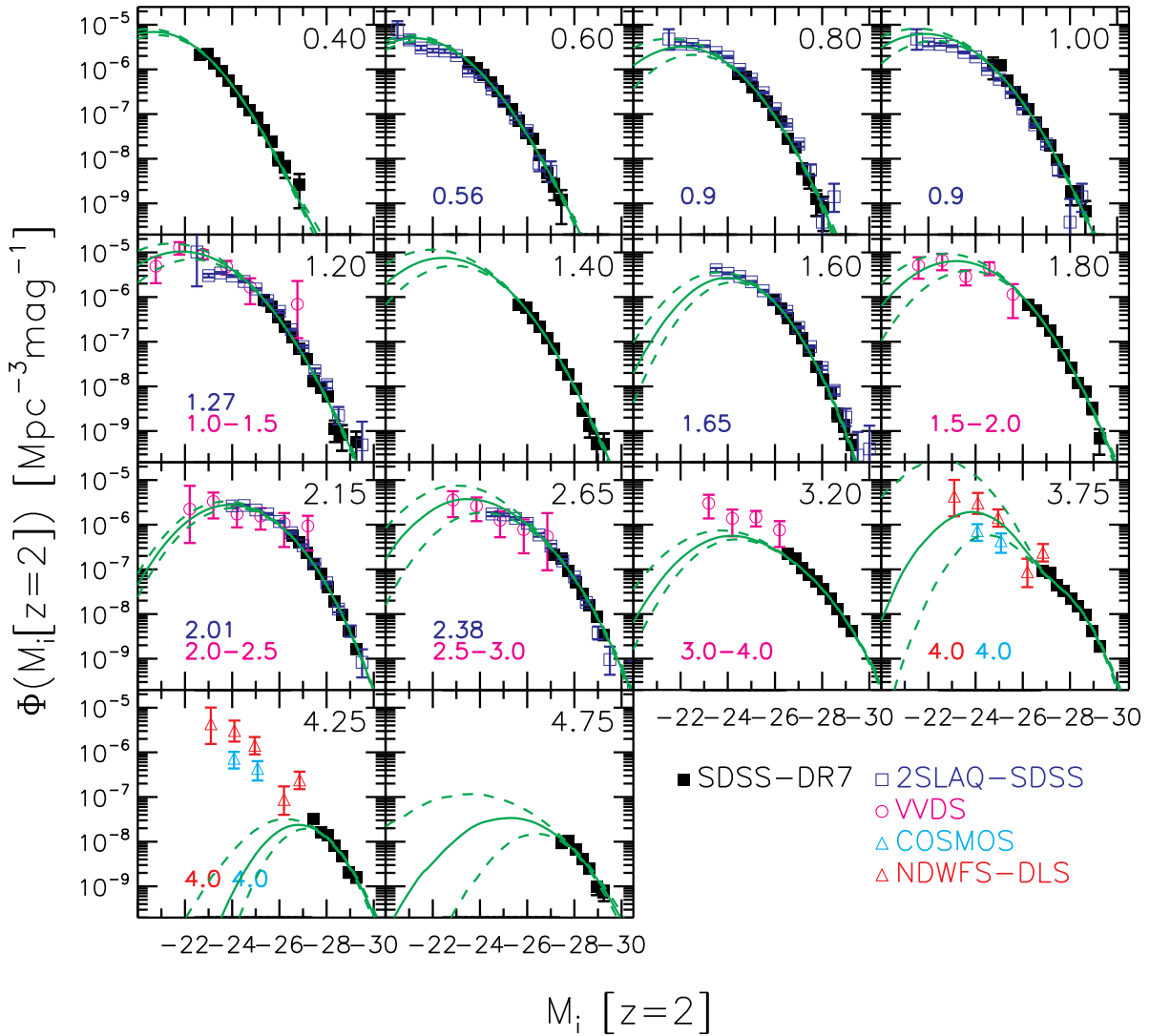}
    \caption{Model LF and comparisons with data. The points represent LF measurements from different surveys:
    SDSS DR7 (black, this work); 2SLAQ-SDSS \citep[blue,][]{Croom_etal_2009}; VVDS \citep[magenta,][]{
	Bongiorno_etal_2007}; COSMOS \citep[cyan,][]{Ikeda_etal_2011}; NDWFS-DLS \citep[red,][]{Glikman_etal_2011}.
    The green solid and dashed lines indicate the median and 68\% percentile of our model LF.
    The constraint in a \texttt{zbin} is generally better if there are more quasars in that bin. The redshift of each
    \texttt{zbin} is marked in the upper-right corner, and the redshifts for other LF data are marked in
    the lower-left corner in the corresponding colors. }
    \label{fig:lf_model}
\end{figure*}

\begin{figure*}
  \centering
    \includegraphics[width=0.9\textwidth]{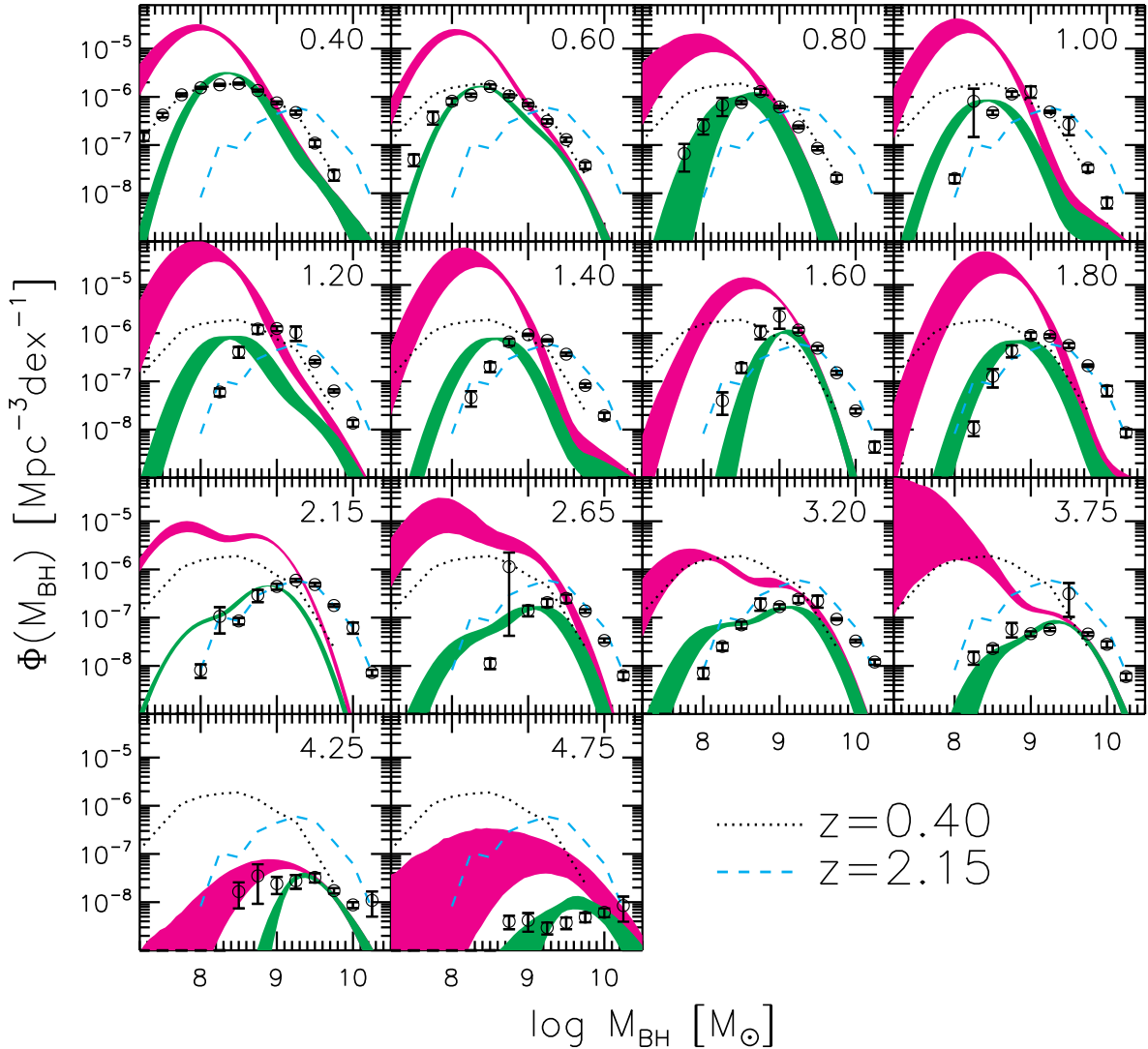}
    \caption{Model BHMF in broad-line quasars. The data points are the binned virial BHMF as in Fig.\
    \ref{fig:vmax_bhmf_dlogm}. The magenta shaded region indicates the 68\%
    percentile range of the model BHMF for all broad-line quasars, while the
    green shaded region indicates that for the detectable quasars (e.g.,
    above the flux limit). The constraint in a \texttt{zbin} is generally
    better if there are more quasars in that bin. Note that in general $d\log M_{\rm BH}\neq d\log M_{\rm BH,vir}$ hence
    the areas underlying the green curve and the data points are not necessarily the same. }
    \label{fig:bhmf_model}
\end{figure*}

\subsection{The LF and the BHMF}

\begin{figure}
  \centering
    \includegraphics[width=0.48\textwidth]{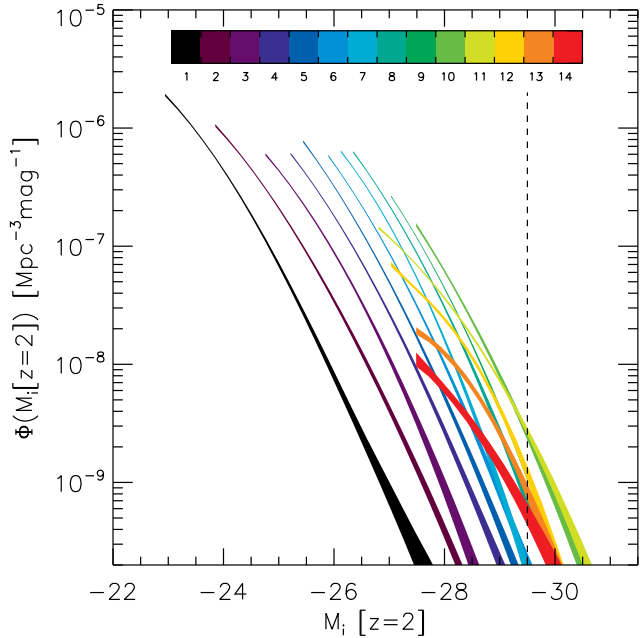}
    \caption{Model LF in the 14 \texttt{zbins}. The width of each stripe indicates the
    $68\%$ range of the model LF. We only show the portion that
    is beyond the flux limit in each \texttt{zbin} and is well constrained by
    the data. The dashed vertical line indicates $M_i(z=2)=-29.5$, beyond which there is no
    binned LF data (see Fig.\ \ref{fig:lf_model}). The slope for $M_i[z=2]\lesssim -27$ is significantly flattened
    beyond \texttt{zbin11} ($\bar{z}=3.2$), as noted in \citet{Richards_etal_2006a}.
    It is also notable that the slope for \texttt{zbin7} is steeper than the adjacent
    two \texttt{zbins}, which is highlighted in Fig.\ \ref{fig:lf_zbin678} and
    discussed in the text.}
    \label{fig:lf_model_all_in_one}
\end{figure}

\begin{figure}
  \centering
    \includegraphics[width=0.48\textwidth]{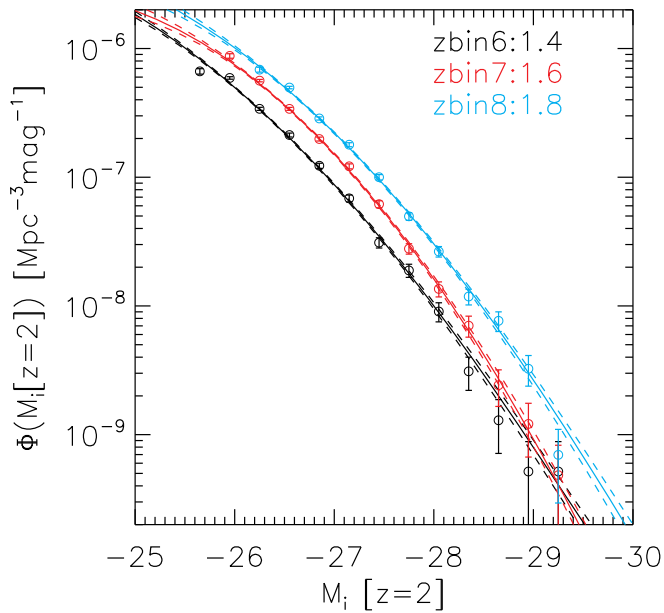}
    \caption{Model LF for \texttt{zbin6,7,8} and comparison to the binned LF (open circles
    with error bars). The solid and dashed lines indicate the median and 68\% range
    of the model LF. The binned LF and our model LF are estimated in completely different
    ways and are in excellent agreement. The LF at $z\sim 1.6$ has a steeper bright-end
    slope than at $z\sim 1.4$ and $z\sim 1.8$, which may reflect systematics in the emission
    line $K$-correction at this redshift (see the text for more discussions).}
    \label{fig:lf_zbin678}
\end{figure}

\begin{figure*}
  \centering
    \includegraphics[width=0.45\textwidth]{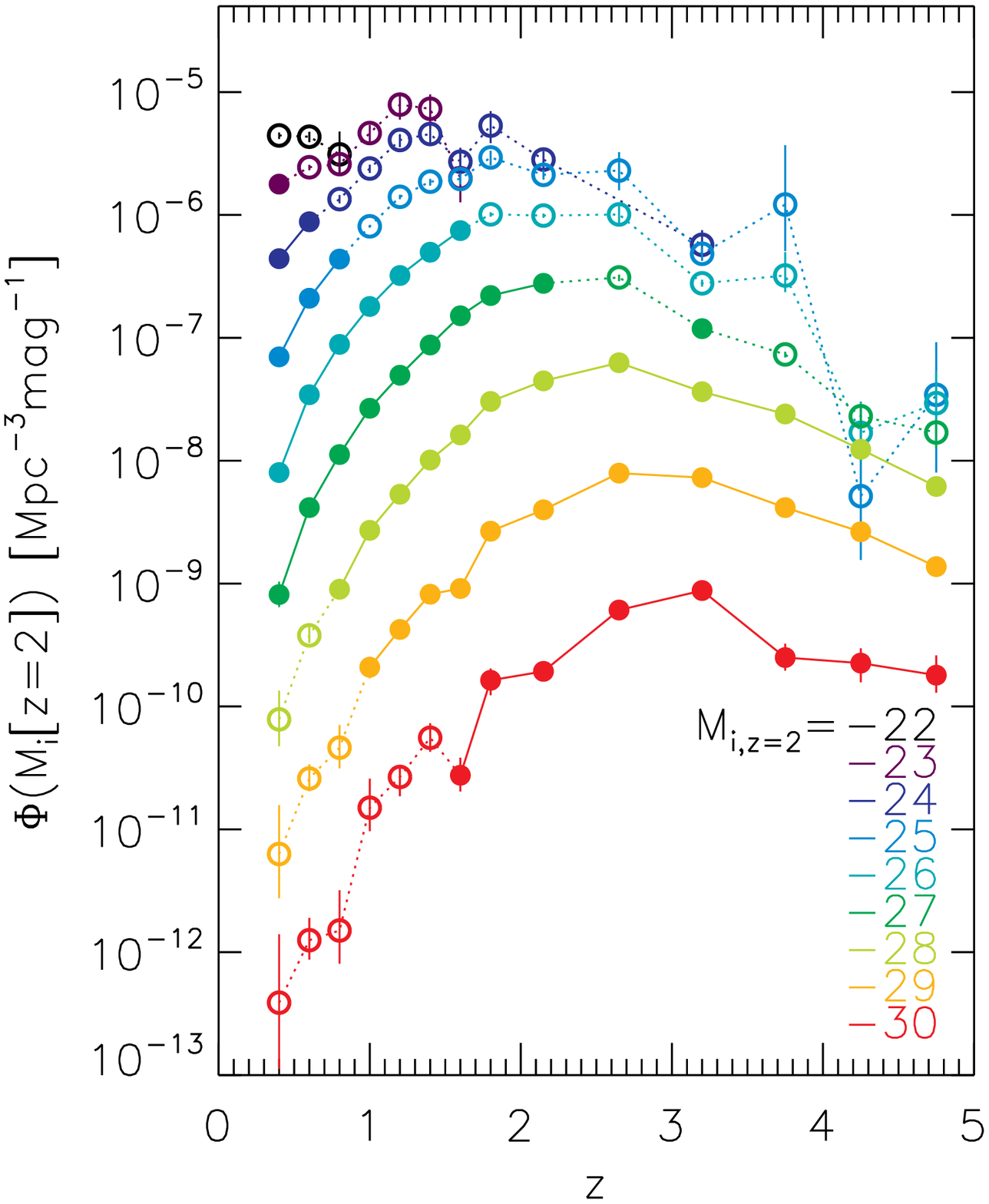}
    \includegraphics[width=0.45\textwidth]{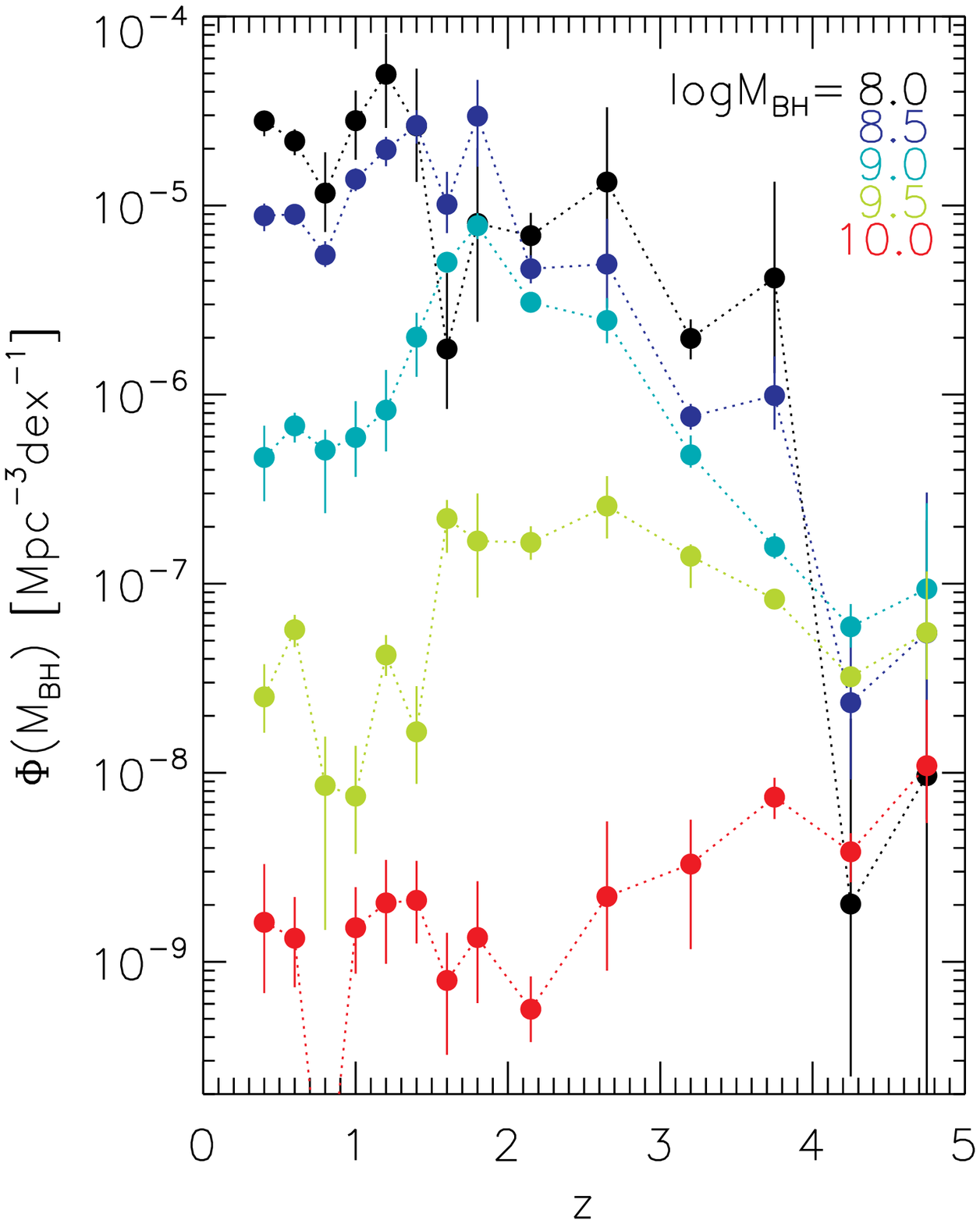}
    \caption{Downsizing of broad-line quasars. {\em Left}: the evolution of quasar number density
    per luminosity interval as a function of luminosity, for our model LF. Filled circles are
    portions of our model LF that are sampled by SDSS quasars, while open circles represent
    model extrapolations. Error bars stand for the 68\% percentile range from our model
    LF. The number density of fainter quasars peaks at later time. The glitch
    at $z=1.6$ (\texttt{zbin7}) is discussed in \S\ref{sec:disc_evo}.
    {\em Right}: the evolution of quasar number density per BH mass interval as a function of
    BH mass, for our model BHMF (all broad-line quasars). Error bars stand for the 68\% percentile
    range from our model BHMF. The estimates are quite noisy due to the
    large uncertainties of our model BHMF. But there is some evidence that the number density of
    lower mass BHs peaks at later time. }
    \label{fig:lf_model_downsizing}
\end{figure*}

\begin{figure}
  \centering
    \includegraphics[width=0.48\textwidth]{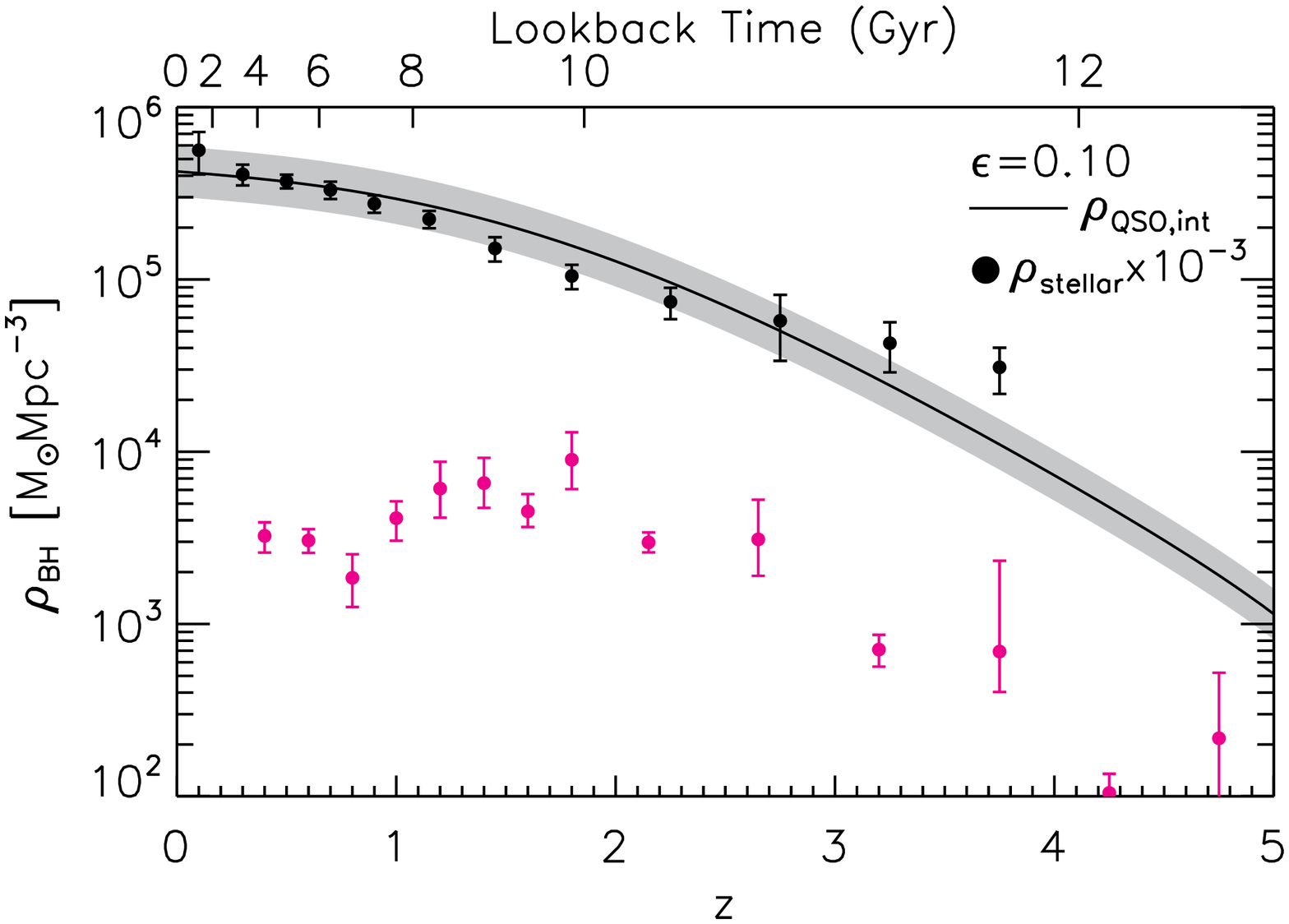}
    \caption{The growth of total BH mass density with cosmic time. The black solid
    line is the accreted BH mass density integrated over the bolometric LF determined by
    \citet{Hopkins_etal_2007a}, assuming a radiative efficiency $\epsilon=0.1$. The gray
    shaded region has a vertical width of $0.3$ dex centered on the solid line, which we
    use as a conservative estimate of the uncertainty of the accreted BH mass density
    due to uncertainties in the bolometric LF and radiative efficiency. The black
    circles are the measured global stellar mass density \citep{Perez-Gonzalez_etal_2008},
    scaled by a factor of $10^{-3}$. The magenta circles are the total BH mass density in
    broad-line quasars, estimated by integrating over our model BHMF. See discussions in
    the text. }
    \label{fig:soltan}
\end{figure}

Given the full posterior distributions of model parameters, we now proceed to
compute the model LF and BHMF using Eqn.\ (\ref{eqn:bhmf})-(\ref{eqn:lf}).
Instead of using the median and percentile of model parameters, we estimate
the median and 68\% percentile of the LF and BHMF from individual LF and BHMF
realizations. We tabulate the model LF and BHMF in Table
\ref{table:phi_m_model}.

Fig.\ \ref{fig:lf_model} compares our model LF with the observed LF. The
black data points are the binned DR7 LF, and other color-coded data points
are from various deeper quasar surveys
\citep[][]{Bongiorno_etal_2007,Croom_etal_2009,Ikeda_etal_2011,
Glikman_etal_2011}. Since the LF data for deeper surveys were not necessarily
measured in exactly the same redshift bin as our grid, we plot them in the
closest bins possible and we indicate their redshifts in the corresponding
colors. The green lines are our model LF and the 68\% percentile range. The
conversions between different magnitudes and $M_i[z=2]$ are given below
\citep[e.g.,][]{Richards_etal_2006a,Croom_etal_2009}:
\begin{eqnarray}
M_i[z=2] & = & M_{1450} - 2.5\alpha_\nu\log\left(\frac{1450\,\textrm{\AA}}{7471\, \textrm{\AA}}\right)-2.5(1+\alpha_\nu)\log 3 \nonumber \\
& = & M_{B}({\rm vega}) - 0.804 \nonumber \\
&=& M_g[z=2] - 2.5\alpha_\nu\log\left(\frac{4670\,\textrm{\AA}}{7471\, \textrm{\AA}}\right) \nonumber \\
&=& -2.5\log\left(\frac{\lambda L}{4\pi c d^2}\right)-48.6-2.5\log 3 \ ,
\end{eqnarray}
where $d=10$ pc, $c$ is the speed of light, $\lambda=2500\,$\AA\, and we
assume a continuum slope $\alpha_\nu=-0.5$.

Our model LF is constrained by the SDSS quasars alone, but it also provides
reasonable prediction when extrapolated to $\sim 3$ magnitudes fainter. This
means our BHMF and Eddington ratio model is reasonable when extrapolating not
too far beyond the regime probed by SDSS quasars. \citet{Schafer_2007} used a
purely statistical technique to extrapolate the \citet{Richards_etal_2006a}
DR3 LF to fainter magnitudes. Compared with his method, our extrapolation is
more physically grounded: it is constrained by a physical model describing
the underlying BHMF and Eddington ratio distribution; although the
mixed-Gaussiaon parametrization for the true BHMF does not have any
particular physical meaning, some model parameters do, such as the mean and
dispersion of Eddington ratios at fixed true mass. There are, however, some
notable discrepancies in the highest redshift bins between our model
extrapolation and the observed faint-end LF. In particular, our model
extrapolation is unable to match the faint-end LF results in
\citet{Glikman_etal_2011} and \citet{Ikeda_etal_2011} at $z\sim 4.25$. This
is most likely caused by the failure of our model extrapolation due to the
much poorer statistics and systematics with the \CIV-based virial masses in
the two highest redshift bins.

Fig.\ \ref{fig:bhmf_model} shows our model BHMF in broad-line quasars. The
magenta shaded region indicates the $68\%$ percentile of the true BHMF, and
the green shaded region indicates the $68\%$ percentile of the portion of
BHMF that can be detected in the flux-limited SDSS sample. The data points
are the binned virial BHMF measured in \S\ref{subsec:bin_BHMF}. The flux
limit of SDSS greatly reduces the abundance of active BHs towards the
low-mass end. At the same time, the shape and peak BH mass are generally
different for the true BHMF and for the observed virial mass function, which
is caused by the difference between true masses and virial masses (see Eqn.\
\ref{eqn:B_corr}). The model BHMF (for all active BHs or detected BHs) has a
much larger uncertainty than indicated by the Poisson errors associated with
the binned virial mass function. This large uncertainty is mostly caused by
the flexibility of our model and the poorly constrained luminosity-dependent
bias, and secondly by the fact that the SDSS quasar sample only probes the
high-luminosity tail of the distribution of quasars.

Several studies have suggested that \CIV\ is the least reliable line to
estimate BH masses
\citep[e.g.,][]{Baskin_Laor_2005,Sulentic_etal_2007,Shen_etal_2008b,Richards_etal_2011},
due to a possible non-virial component that contributes to the line profile.
The model constraints are particularly poor in the highest redshift bins. Our
rather simplistic model for the relation between virial masses and true
masses may not work very well for the problematic \CIV\ estimator. A
re-assessment and possible improvement of the current version of the \CIV\
estimator is desirable \citep[e.g.,][Shen et~al., in
preparation]{Assef_etal_2011}.

We can obtain tighter constraints on the BHMF and model parameters if we
impose more restrictive prior constraints. For instance, if we fix the value
of $\beta$, the uncertainties of the resulting BHMF and model parameters are
substantially reduced, while the LF is not changed significantly. This
suggests that better prior knowledge on these parameters can help improve
these model constraints significantly.

\section{Discussion}\label{sec:disc}

\subsection{Evolution of quasar demographics}\label{sec:disc_evo}

It is well know that the number density of bright quasars peaks around
redshift $\sim 2-3$ \citep[e.g.,][]{Richards_etal_2006a}. The most important
result in quasar demographics in the past decade is the so-called ``cosmic
downsizing'', i.e., the number density of less luminous objects peaks at
lower redshift. Initially discovered in the X-ray surveys
\citep[e.g.,][]{Cowie_etal_2003,Steffen_etal_2003,Ueda_etal_2003,Hasinger_etal_2005},
this trend is now confirmed in the optical band as well
\citep[e.g.,][]{Bongiorno_etal_2007,Croom_etal_2009}.

Here we use the DR7 quasar sample and our model LF/BHMF to probe downsizing
in quasar demographics. The flux limit of SDSS quasars is generally not deep
enough to probe the evolution of the number density of the less luminous
quasars. However, we have the best constraints on the high-luminosity end,
and our model LF can extrapolate down to fainter luminosities, thus
compensating for the shallow depth of the SDSS quasar sample.

Fig.\ \ref{fig:lf_model_all_in_one} shows the model LF in all 14
\texttt{zbins}. We only plot the portion that is above the flux limit in each
individual bin for which our model LF reproduces the binned LF well and the
constraints are the tightest. The width of each stripe indicates the
1$\sigma$ ($68\%$) range of the model LF. The dashed vertical line indicates
$M_i(z=2)=-29.5$, beyond which there is no binned LF data (see Fig.\
\ref{fig:lf_model}). The most noticeable feature is that the curvature of the
LF changes significantly beyond $z\sim 3$. This is noted in
\citet{Richards_etal_2006a} as the flattening of the bright-end slope, if a
single power-law were fitted to the binned LF data. However, it appears to be
more of a strong evolution of the break luminosity than a change in the
bright-end slope. Recently, \citet{Fontanot_etal_2007} argue that the
apparent flattening in the bright-end slope is caused by an overestimation of
the completeness of SDSS color selection at $z>3.5$, as was adopted in
\citet{Richards_etal_2006a}. An implicit assumption in their argument is that
the completeness must become lower towards the flux limit, leading to an
artificial flattening in the bright-end slope. However, we found that in our
model LF, the break luminosity in the LF gradually increases towards higher
redshift in a continuous fashion (see Fig.\ \ref{fig:lf_model_all_in_one}),
and the slope is already significantly flattened at $z\sim 3.2$ for
$M_i[z=2]\la -27$. Thus it is not obvious that this flattening in the
bright-end slope is caused by the possible differential incompleteness in the
SDSS color selection. A more careful analysis is needed to resolve this
issue.

Another notable feature is that \texttt{zbin7} ($z=1.6$) shows a steeper
slope at $M_i[z=2]<-28$ than the adjacent \texttt{zbins}. This is not caused
by the failure of our model, as shown in Fig.\ \ref{fig:lf_model}. A close
comparison between our model LF and the binned LF is shown in Fig.\
\ref{fig:lf_zbin678} for \texttt{zbin6}, \texttt{zbin7} and \texttt{zbin8}.
The points show the binned LF data, and the lines show the model LF from our
Bayesian approach; the two are in excellent agreement. This apparent
deviation in \texttt{zbin7} is likely caused by the systematics of emission
line $K$-correction. For this \texttt{zbin}, the \MgII\ line contributes the
most to $i$-band. The \MgII\ line strength relative to the continuum
decreases with luminosity \citep[e.g., the Baldwin effect, Baldwin 1977; see
fig.\ 13 in][]{Shen_etal_2011a}, thus using a constant emission line
$K$-correction in this bin will lead to more underestimated continuum
luminosity towards the brighter end. This causes an artificial steepening at
the bright end of the \texttt{zbin7} LF (a similar effect is seen for
\texttt{zbin12} where \CIV\ contributes the most to the $i$-band). This is a
second-order effect and therefore we do not correct for it in the present
work. However, it does indicate that as the statistics becomes much better,
additional systematic effects need to be taken into account in order to get
unbiased results. We note that \texttt{zbin7} also shows notable deviations
from \texttt{zbin6} and \texttt{zbin8} in terms of the extrapolated LF and
constraints on model parameters (see Figs.\ \ref{fig:model_evo} and
\ref{fig:lf_model}).

Finally, as mentioned earlier, the LF (above the SDSS flux limit) generally
flattens (i.e., the luminosity break) at brighter magnitude towards higher
redshift, consistent with earlier findings \citep[e.g.,][]{Croom_etal_2009}.
However, we did not observe a significant steepening of the bright-end slope
from $z=0.4$ to $z=2.65$, in tension with \citet{Croom_etal_2009}. We suspect
this discrepancy is caused by the different methods to measure the LF:
\citet{Croom_etal_2009} fit the LF with a broken power-law function, while
our LF estimate is non-parametric.

Fig.\ \ref{fig:lf_model_downsizing} ({\em left}) shows the evolution of
quasar number density as a function of luminosity, using our model LF. Filled
circles indicate the portion of the LF that is sampled by SDSS quasars, and
open circles indicates extrapolated model LF. Note that at low redshift, the
SDSS sample also suffers from the bright $i=15$ cut, in addition to the faint
luminosity cut. We only extrapolate the model LF to 3 magnitudes fainter,
where our model constraints are still reasonably good. With the addition of
extrapolated data, there is a clear trend that the number density of more
luminous quasars peak earlier. Compared with the results from 2SLAQ+SDSS
\citep[][]{Croom_etal_2009}, our sample extends to higher redshift and the
peaks for the bright quasars are well resolved.

Fig.\ \ref{fig:lf_model_downsizing} ({\em right}) shows the evolution of
quasar number density as a function of BH mass. The uncertainties in each
redshift bin are quite large, reflecting the poorly constrained BHMF (e.g.,
Fig.\ \ref{fig:bhmf_model}). However, there is tentative evidence that more
massive BHs achieve their peak density earlier. This is the manifestation of
``cosmic downsizing'' in terms of BH mass. As redshift decreases, active BHs
become on average less massive and are likely in less massive hosts. The
number density of $M_{\rm BH}\sim 10^9\,M_\odot$ quasars peaks around $z=2$,
which is consistent with the trend found in \S\ref{subsec:bin_BHMF} based on
virial masses (although the characteristic BH mass shifted to $\sim 3\times
10^9\, M_\odot$ for virial masses due to the luminosity-dependent bias in the
flux limited sample). \citet{Kelly_etal_2010} also found evidence for active
BH mass downsizing, and \citet{Vestergaard_Osmer_2009} found downsizing at
the high-mass end of the active BHMF, although the latter did not correct for
incompleteness and scatter in virial mass estimates.

We can compare the derived active BHMF with the local dormant BHMF estimated
by \citet{Shankar_etal_2009a}. Below $\sim 3\times 10^9\,M_\odot$ (the upper
limit that can be measured for the local BHMF), the active BHMF at any
redshift is always more than one order of magnitude lower than the local
dormant BHMF. This means quasars are by no means a long-lived cosmological
population, and we are witnessing different generations of quasars at
different redshift, which cumulatively (over time) build up the local dormant
BH population.

Fig.\ \ref{fig:soltan} further demonstrates the insignificant contribution of
broad-line quasars to the total BH mass density at all redshifts. The black
line shows the accreted BH mass density using the bolometric LF estimated by
\citet{Hopkins_etal_2007a} and a radiative efficiency $\epsilon=0.1$, where
the gray shaded region is centered on the black line and has a vertical width
of $0.3$ dex, which is a conservative estimate for the uncertainty in the
accreted BH mass density due to uncertainties in the bolometric LF and
radiative efficiency. The accreted BH mass density at $z=0$ is consistent
with the estimated BH mass density using local spheroid-BH scaling relations
\citep[e.g.,][]{Shankar_etal_2009a,Yu_Lu_2008}. The black circles are the
measured global stellar mass density in \citet{Perez-Gonzalez_etal_2008},
scaled by a factor of $10^{-3}$. Although there are systematic uncertainties
in both the accreted BH mass density and the measured global stellar mass
density, the agreement between the two is remarkable, and argues strongly for
the co-evolution between galaxies and BHs. However, the total BH mass density
in broad-line quasars (magenta circles), estimated by integrating over our
model BHMF, is at least one order of magnitude less than the total BH mass
density at all times.

\subsection{The Eddington ratio distribution}\label{sec:disc_edd}

\begin{figure}
  \centering
    \includegraphics[width=0.48\textwidth]{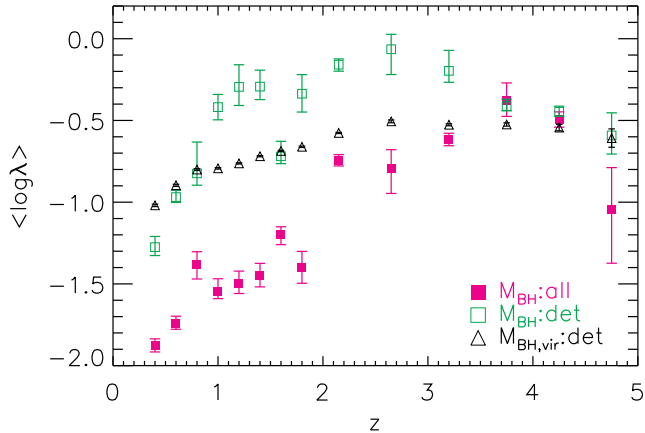}
    \caption{Evolution of the sample-averaged Eddington ratio. The magenta points
    are for all broad-line quasars with true masses; the green points are for the
    detectable quasars with true masses; both are estimated from our model. The black points
    are the observed average Eddington ratio in our quasar sample, estimated using
    virial BH masses. Note that the error bars here indicate the uncertainty in the mean
    value, not the scatter in Eddington ratios. }
    \label{fig:eddington_dist}
\end{figure}

Our model specifies the Eddington ratio distribution (or luminosity
distribution) at fixed BH mass\footnote{The Eddington ratio distribution at
fixed BH mass is different from that at fixed luminosity, in the presence of
scatter of the $M_{\rm BH}-L_{\rm bol}$ relation (cf. Fig.\
\ref{fig:zbin2_ml_plane_all}). This is true no matter which mass is used
(virial versus true). For the virial mass-based Eddington ratio at fixed
luminosity, the average value scales as $L^{0.5}\propto M_{\rm BH,vir}$ from
the virial relation (assuming a slope of $0.5$ in the mean $R-L$ relation).
However, for the virial mass-based Eddington ratio at fixed virial mass, the
average value has a much weaker dependence on $M_{\rm BH,vir}$, and one also
must account for the selection effect of the flux limit. These points are
well demonstrated in Fig.\ \ref{fig:zbin2_ml_plane_all}.}. Even though the
constrained parameters have large uncertainties (see Fig.\
\ref{fig:model_evo}), it appears that a linear relation between the mean
luminosity and mass is consistent with the data. This means that the mean
Eddington ratio does not depend on mass strongly. However, there is some
evidence that the mean Eddington ratio increases with redshift, which may
reflect the evolution of the average accretion efficiency or the general
decline of quasar light curve. The scatter in the Eddington ratio
distribution is poorly constrained, and no coherent redshift evolution is
seen. Nevertheless, we found a median value of $\sim 0.4$ dex in the
dispersion of Eddington ratios at fixed BH mass, consistent with earlier
studies \citep[e.g.,][]{Shen_etal_2008b,Kelly_etal_2010}.

The Eddington ratio distribution in any flux-limited sample or at fixed
luminosity is essentially different from that of all broad-line quasars or at
fixed BH mass, as emphasized in \citet{Shen_etal_2008b} and
\citet{Kelly_etal_2010}. It suffers from the Malmquist-type bias such that
more high-Eddington ratio and low mass objects are scattered into the
flux-limited sample than low-Eddington ratio and high mass objects scattered
out (i.e., due to a non-zero $\sigma_l$ and a bottom-heavy BHMF). Thus the
average Eddington ratio is biased high in our sample. In addition, using
virial masses can further modify the sample-averaged Eddington ratio due to
the luminosity-dependent bias (i.e., a non-zero $\beta$).

Fig.\ \ref{fig:eddington_dist} summarizes these behaviors. Plotted here are
the sample-averaged Eddington ratio and its uncertainty\footnote{Note that
the uncertainty here refers to the uncertainty in the mean Eddington ratio,
not the scatter in Eddington ratio.} in 14 \texttt{zbins} under different
circumstances. The filled magenta squares are for all the broad-line quasars,
and the open green squares are for those that are detectable above the flux
limit; both are estimated using our model outputs. The difference between the
mean Eddington ratio for all quasars and detectable quasars reflects the
scatter in luminosity at fixed true mass ($\sigma_l$) and the shape of the
underlying BHMF, and in general the mean Eddington ratio in the detectable
sample is biased high. The open triangles are for all quasars in our sample,
where the Eddington ratios are estimated with virial BH masses. The
difference between the mean Eddington ratio for the detectable quasars
estimated with true masses and virial masses mostly reflects the
luminosity-dependent bias (i.e., a non-zero $\beta$). Since we have a
positive $\beta$ for most \texttt{zbins}, the mean Eddington ratio based on
virial masses for the detectable quasars is generally biased low than the
true value\footnote{When $\beta$ is small, the difference between the mean
Eddington ratio with true masses and virial masses for the flux-limited
sample may not be discernable, due to the slightly different calculations of
the mean Eddington ratio in the two cases: the mean {\em true} Eddington
ratio is calculated using random draws from the posterior distribution; the
mean {\em virial} Eddington ratio is the {\em median} in the observed
distribution. }.

\subsection{Model caveats and future perspectives}\label{sec:disc_caveats}

Our forward Bayesian modeling stands for a major improvement on retrieving
information from the mass-luminosity plane compared with previous studies
based on virial BH masses. However, there are still several caveats in our
approach, which can be improved in future investigations:

\begin{itemize}

\item First and foremost, we assumed that the specific forms of
    single-epoch virial mass estimators adopted in this work give
    unbiased BH mass estimates when averaged over luminosity at fixed
    true mass, i.e., $E(m_e|m)=m$, where $E(...)$ stands for the
    expectation value. While this is true for the local RM AGN sample (by
    calibration), it may not hold when extrapolating to the high-redshift
    and high-luminosity population.

    Also, we recognize that different versions of virial estimators (even
    for the same line) do not necessarily yield the same BH masses. If
    the virial mass estimators that we used were already biased for
    objects in our sample (i.e., due to an imperfect virial calibration,
    or other generic systematics in the virial technique) then the BHMF
    estimation would also be biased. In particular, it is possible that
    FWHM is a worse indicator for the virial velocity than other line
    width measures \citep[such as line dispersion, see arguments
    by][]{Peterson_etal_2004,Collin_etal_2006,Wang_etal_2009b,Rafiee_Hall_2011a,Rafiee_Hall_2011b},
    and the luminosity bias we found here may reflect the systemics with
    incorrect forms of virial estimators.

    Nevertheless, there is currently no consensus on which versions of
    virial estimator are the best. As a proof of concept, our model can
    easily be adapted to use the updated virial calibrations when they
    become available. A similar study based on line dispersion-based
    virial BH masses is currently under way.

\item Our parameterized model is still in a relatively restrictive form,
    especially for the the Eddington ratio model at fixed true mass. We
    have tested with rather relaxed parameterizations for the joint
    distribution of luminosity and true mass, but found that the current
    data is not sufficient to give reasonable constraints if we treat the
    luminosity-dependent bias as a free parameter. This can be seen in
    Fig.\ \ref{fig:zbin2_ml_plane_all} that the current data only sample
    the tip of the distribution, and thus it is difficult to constrain
    parameters for overly flexible models. Deeper data is needed to test
    more flexible models. A fainter broad-line quasar sample, such as the
    2SLAQ sample \citep{Croom_etal_2009} or the BOSS quasar sample
    \citep{Ross_etal_2011}, will also provide more stringent constraints
    on our model parameters and better representation of the
    mass-luminosity plane.

    Nevertheless, the capability of reproducing the observed distribution
    is reassuring that a single lognormal Eddington ratio model with a
    mass-dependent mean is still a reasonable choice. An alternative
    model would be a power-law Eddington ratio distribution at fixed true
    mass, as suggested in some models for the decaying part of the quasar
    light curve \citep[e.g.,][]{Yu_Lu_2004,Hopkins_etal_2005a,Shen_2009}.
    However, since quasars are continuously forming (especially for
    fainter quasars, whose number density peaks later in time), at each
    epoch we should also witness quasars in their rising part of the
    light curve, unless this period is completely obscured. In addition,
    the {\em broad line region} may only exist when the BH is accreting
    at high accretion rate during AGN evolution
    \citep[e.g.,][]{Shen_etal_2007a,Hopkins_etal_2009}, i.e., with
    Eddington ratio $\lambda\sim 10^{-3}- 1$
    \citep[e.g.,][]{Trump_etal_2011}, and there must be fluctuations in
    the instantaneous Eddington ratio. Thus a lognormal model may indeed
    capture the basic characteristic of the Eddington ratio distribution
    at fixe BH mass.

    In a follow-up paper (Kelly \& Shen, in preparation), we will use a
    more generic prescription to model the joint distribution of
    luminosity and BH mass directly instead of using a specific Eddington
    ratio model as adopted here. This will be achieved at the cost of
    simplifying other prescriptions in the model, for instance, fixing
    the value of $\beta$. However, by relaxing the single lognormal
    Eddington ratio model, we can test if a single lognormal model is a
    good approximation. It also allows a more proper investigation of the
    mass-luminosity plane.

    Finally, we note that we assumed a Gaussian distribution for the
    error in the virial mass estimates (which may not be adequate), and
    our model is not robust to outliers. These issues may be partly
    responsible for the large uncertainties we get in several
    \texttt{zbins}. We plan to investigate these issues in future work.

\item Since we are only concerned with broad-line objects, the active
    BHMF derived in this paper is a lower-limit for the active SMBH
    population. It is important to include the growth of SMBHs in
    obscured accretion or with accretion mode that does not produce broad
    lines, as traced by populations of active BHs selected by various
    other techniques
    \citep[e.g.,][]{Stern_etal_2005,Reyes_etal_2008,Luo_etal_2008,
    Treister_etal_2009,Hickox_etal_2009,Elvis_etal_2009,Yan_etal_2011,Xue_etal_2011,Civano_etal_2011}.
    This non-broad-line population could contribute as much as one half
    of the total accreted BH mass density, although significant
    uncertainty remains in determining their relative abundance to the
    broad-line population.


\end{itemize}

\section{Conclusions}\label{sec:con}

This paper represents our first step towards a systematic investigation on
the demographics of broad-line quasars in the mass-luminosity plane and its
redshift evolution. We used a forward modeling Bayesian framework to model
the joint distribution in the mass-luminosity plane. With simple model
prescriptions for the underlying active BHMF and Eddington ratio
distribution, we were able to fit the observed distribution above the sample
flux limit and extrapolate below using constrained model parameters. We paid
particular attention to the distinction between virial mass estimates and
true masses, and corrected for the selection effects of flux limited samples.
The main conclusions of the paper are the following:

\begin{enumerate}

\item[1.] Virial BH masses are not true masses (\S\ref{subsec:prelim}),
    and their explicit dependence on luminosity has implications for
    their conditional probability distributions: $p(m_e|m)\neq
    p(m_e|m,l)\neq p(m_e|l)$, and $\sigma_{\rm vir}>\sigma_{lm}$. We
    found evidence that for \MgII\ and possibly \CIV, there is a positive
    luminosity-dependent bias in virial masses, such that at fixed true
    mass, objects with luminosities above average have over-estimated
    virial masses. This is expected as there must be uncorrelated scatter
    between line width and luminosity at fixed true mass, due to the
    imperfectness of them being the surrogates for the virial velocity
    and the BLR radius in the virial mass technique. While this bias was
    noted earlier \citep{Shen_etal_2008b,Shen_Kelly_2010}, this is for
    the first time that we quantified the level of this bias with the
    SDSS quasar sample. As a consequence, the dispersion in virial mass
    in narrow luminosity bins or flux-limited samples is generally
    smaller than the virial mass uncertainty
    \citep{Shen_etal_2008b,Shen_Kelly_2010}, and the average virial BH
    masses in $z\ga 0.7$ SDSS quasars are likely overestimated by a
    factor of a few.

\item[2.] Our model LF is tightly constrained in the regime sampled by
    SDSS quasars, and makes reasonable predictions when extrapolated
    $\sim 3$ magnitudes fainter, as compared with the LF measured from
    faint quasar surveys. Downsizing in LF evolution is clearly seen with
    our model LF. The break luminosity (as in a double power-law model)
    increases towards higher redshift. As a consequence, the LF slope for
    $M_i[z=2]<-27$ from a single power-law fit appears shallower at $z>3$
    than at lower redshift.

\item[3.] Except for a few \texttt{zbins}, the model active BHMF usually
    cannot be well constrained to within a factor of a few. This is
    mainly caused by the additional free parameter on the
    luminosity-dependent bias, which leads to degeneracies with other
    parameters. The SDSS sample only probes the tip of the active SMBH
    population at high redshift, which makes it more difficult to
    constrain model parameters for the whole population. Nevertheless,
    the abundance of the most massive quasars is always overestimated
    using virial masses regardless of the luminosity-dependent bias.
    Within our model a turn-over at the low-mass end of the true BHMF is
    needed in order to fit the observed distribution in the
    mass-luminosity plane, but we caution that this result is based on
    our model assumptions and may suffer from the poor sampling of
    low-mass BHs in the SDSS sample. Further investigations are needed to
    test the robustness of this feature. This turn-over shifts to larger
    BH masses for the detected population in SDSS due to the flux limit.
    Although with large error bars, we found tentative evidence that
    downsizing also manifests itself in the active BHMF, and the number
    density of $M_{\rm BH}\sim 10^9\,M_\odot$ quasars peaks around $z\sim
    2$.

    The total BH mass density in broad-line quasars is always
    insignificant compared with that in all SMBHs at any redshift (Fig.\
    \ref{fig:soltan}).

\item[4.] Within our model uncertainties, the Eddington ratio
    distribution at fixed true BH mass has a mean value that weakly
    depends on mass, and an average scatter of $\sim 0.4$ dex. The
    sample-averaged Eddington ratio for all broad-line quasars ranges
    between $\sim 0.01$ and $\sim 0.3$, and increases with redshift
    (Fig.\ \ref{fig:eddington_dist}). The sample-averaged Eddington ratio
    for quasars above the SDSS flux limit is Malmquist-biased due to the
    scatter in Eddington ratios at fixed BH mass and the shape of the
    underlying active BHMF. Furthermore, using virial masses for the
    detected quasars tend to reduce the Eddington ratio again due to the
    luminosity-dependent bias (Fig.\ \ref{fig:eddington_dist}).

\item[5.] Our model reproduces the observed distribution in the
    mass-luminosity plane (with virial masses). The flux limit, the
    scatter between true masses and virial masses, as well as the
    luminosity-dependent bias, change the distribution in the
    mass-luminosity plane significantly. Thus any features in the
    mass-luminosity plane based on virial masses must be interpreted with
    caution.

\end{enumerate}

Our results highlight the need for understanding the systematics in the
virial mass technique. All single-epoch virial mass estimators currently
utilized in the literature are bootstrapped from the local RM AGN sample of
only several dozen objects. These mass estimators have been extensively used
for high redshift and high luminosity quasars, a regime that is poorly
sampled by RM AGNs. We have shown that even without other systematic issues
with the virial technique, the scatter and luminosity-dependent bias between
virial masses and true masses already make it difficult to constrain the
active BHMF accurately. Any additional systematic issues will simply make the
situation worse.

Since BH mass is a crucial physical quantity and is related to many
fundamental processes of SMBHs, refining the techniques that weigh the BH is
of tremendous importance. As the most promising method to measure
active BH masses, the RM technique (and its extensions of single-epoch virial
estimators) should be tested and calibrated carefully. Progress is being made
in reverberation mapping studies of broad-line AGNs and calibrations of
virial mass estimators
\citep[e.g.,][]{Kaspi_etal_2007,Bentz_etal_2009b,Woo_etal_2010,Graham_etal_2011},
as well as in the statistical description of AGN variability as applied to
reverberation mapping
\citep[e.g.,][]{Zu_etal_2011,Brewer_etal_2011,Pancoast_etal_2011}. However,
to fully understand the BLR kinematics/geometry and to construct reliable
mass estimators, a substantially larger RM sample is needed to sample an
unbiased parameter space of broad-line objects, as well as a much better
understanding of the systematics in the RM technique and single-epoch mass
estimators.

\acknowledgements

We thank the referee for suggestions that led to improvement of the
manuscript, and Brad Peterson, Gordon Richards, Marianne Vestergaard, Youjun
Lu, Qingjuan Yu, and Francesco Shankar for useful comments on the draft. We
also acknowledge Michael Strauss, Xin Liu, Scott Tremaine and Subo Dong for
helpful discussions during early stages of this work. YS acknowledges support
from a Clay Postdoctoral Fellowship through the Smithsonian Astrophysical
Observatory (SAO). BK acknowledges support by NASA through Hubble Fellowship
grant \#HF-51243.01 awarded by the Space Telescope Science Institute, which
is operated by the Association of Universities for Research in Astronomy,
Inc., for NASA, under contract NAS 5-26555.

Funding for the SDSS and SDSS-II has been provided by the Alfred P. Sloan
Foundation, the Participating Institutions, the National Science Foundation,
the U.S. Department of Energy, the National Aeronautics and Space
Administration, the Japanese Monbukagakusho, the Max Planck Society, and the
Higher Education Funding Council for England. The SDSS Web Site is
http://www.sdss.org/.

Facilities: Sloan

\nocite{*}
\bibliographystyle{apj}

\end{CJK*}
\end{document}